\documentclass[useAMS,usenatbib]{mn2e}
\usepackage{psfig} 
\usepackage{natbib} 
\usepackage{subfigure} 
\usepackage{ulem}

\newcommand{\be}{\begin{equation}}
\newcommand{\ee}{\end{equation}}
\newcommand{\ba}{\begin{eqnarray}}
\newcommand{\ea}{\end{eqnarray}}
\newcommand{\brr}{\begin{array}}
\newcommand{\err}{\end{array}}
\newcommand{\bc}{\begin{center}}
\newcommand{\ec}{\end{center}}
\newcommand{\hm}{\,h^{-1}{\rm Mpc}}

\newcommand{\msun}{\,h^{-1}M_\odot}

\newcommand{\vel}{\,{\rm km\,s^{-1}}}

\newcommand{\gadget}{{\footnotesize {\sc GADGET~}}}

\newcommand{\nr}{{\footnotesize {\sc NR}}}
\newcommand{\nrsv}{{\footnotesize {\sc NR-SV}}}
\newcommand{\nrrv}{{\footnotesize {\sc NR-RV}}}
\newcommand{\csf}{{\footnotesize {\sc CSF}}}
\newcommand{\csfc}{{\footnotesize {\sc CSF-C}}}
\newcommand{\w}{{\footnotesize {\sc CSF-M-W}}}
\newcommand{\nw}{{\footnotesize {\sc CSF-M-NW}}}
\newcommand{\agn}{{\footnotesize {\sc CSF-M-AGN}}}
\pssilent 

\newcommand{\mg}{\mbox{$M_{\rm{gas}}$}}
\newcommand{\tmw}{\mbox{$T_{\rm{mw}}$}}
\newcommand{\yx}{\mbox{$Y_{\rm{X}}$}}
\newcommand{\mmg}{\mbox{$M_{\rm{500}}-M_{\rm{gas}}$}}
\newcommand{\mt}{\mbox{$M_{\rm{500}}-T_{\rm{mw}}$}}
\newcommand{\myx}{\mbox{$M_{\rm{500}}-Y_{\rm{X}}$}}

\newcommand{\mincir}{\raise
  -2.truept\hbox{\rlap{\hbox{$\sim$}}\raise5.truept \hbox{$<$}\ }}
\newcommand{\magcir}{\raise
  -2.truept\hbox{\rlap{\hbox{$\sim$}}\raise5.truept \hbox{$>$}\ }}
\newcommand{\siml}{\raise
  -2.truept\hbox{\rlap{\hbox{$\sim$}}\raise5.truept \hbox{$<$}\ }}
\newcommand{\simg}{\raise
  -2.truept\hbox{\rlap{\hbox{$\sim$}}\raise5.truept \hbox{$>$}\ }}

\title[Mass proxies from simulations] {X-ray mass proxies from
  hydrodynamic simulations of galaxy clusters (paper I)}
\author[Fabjan et al.]
{D. Fabjan$^{1,2,3}$, S. Borgani$^{2,4,5}$, E. Rasia$^{6}$, 
A. Bonafede$^{7}$, K. Dolag$^{8,9}$, G. Murante$^{10,4}$ 
\newauthor \& L. Tornatore$^{2,4,5}$
\\~\\
$^1$ Center of Excellence SPACE-SI, A\v{s}ker\v{c}eva 12, 
1000 Ljubljana, Slovenia (dunja.fabjan@space.si)\\
$^2$ Dipartimento di Fisica dell'Universit\`a di
  Trieste, Sezione di Astronomia, via Tiepolo 11, I-34131 Trieste,
  Italy\\ (borgani,tornatore@oats.inaf.it)\\
$^3$ Faculty of Mathematics and Physics, University of Ljubljana, 
  Jadranska 19, 1000 Ljubljana, Slovenia\\
$^4$ INAF -- Osservatorio Astronomico di Trieste, via
  Tiepolo 11, I-34131 Trieste, Italy\\
$^5$ INFN -- Istituto Nazionale di Fisica Nucleare, Trieste, Italy\\
$^6$ Department of Astronomy, University of Michigan, 500
  Church St., Ann Arbor, MI 48109, USA (rasia@umich.edu)\\
$^7$ Jacobs University Bremen, Campus Ring 1, 28759,
  Bremen, Germany (a.bonafede@jacobs-university.de)\\
$^8$ Universit\"ats-Sternwarte M\"unchen, Scheinerstr. 1, 
  D-81679 M\"unchen, Germany\\
$^9$ Max-Planck-Institut f\"ur Astrophysik, POBox 1317,
  85741, Garching bei M\"unchen, Germany (kdolag@mpa-garching.mpg.de)\\
$^{10}$ INAF --
  Osservatorio Astronomico di Torino, Str. Osservatorio 25,
  I-10025, Pino Torinese, Torino, Italy (murante@oato.inaf.it)
}

\date{Accepted 2011 February 07. Received 2010 December 20, in original form ...}

\begin{document}   

\maketitle
\label{firstpage}

\begin{abstract}
  Using extended sets of cosmological hydrodynamical simulations of
  galaxy clusters, we present a detailed study of scaling relations
  between total cluster mass and three mass proxies based on X--ray
  observable quantities: temperature of the intra--cluster medium, gas
  mass and the product of the two, $Y_X=M_{gas}T$. Our analysis is
  based on two sets of high--resolution hydrodynamical simulations
  performed with the TreePM--SPH \gadget code. The first set includes
  about 140 cluster with masses above $5\times 10^{13}\msun$, with 30 of
  such clusters having mass above $10^{15}\msun$. All such clusters
  have been simulated in two flavours, with both non--radiative
  physics and including cooling, star formation, chemical enrichment
  and the effect of supernova feedback triggering galactic ejecta. The
  large statistics offered by this set of simulated clusters is used to
  quantify the robustness of the scaling relations between mass
  proxies and total mass, to determine their redshift evolution, and
  to calibrate their intrinsic scatter and its
  distribution. Furthermore, we use a smaller set of clusters,
  including 18 halos with masses above $5\times 10^{13}\msun$, four of
  which are more massive than $10^{15}\msun$, to test the robustness
  of mass proxies against changing the physical processes that are included in
  the simulations to describe the evolution of the intra--cluster
  medium. Each cluster is simulated in seven different flavours to study
  the effect of {\it (i)} thermal conduction, {\it (ii)} artificial
  viscosity, {\it (iii)} cooling and star formation, {\it (iv)}
  galactic winds, and {\it (v)} AGN feedback.

  As a general result, we find the $M$--\yx\ scaling relation to be
  the least sensitive one to variations of the ICM physics, its slope
  and redshift evolution being always very close to the predictions of
  the self--similar model. As for the scatter around the best--fitting
  relations, its distribution is always close to a log-normal
  one. \mg\ is the mass proxy having the smallest scatter in mass,
  with values of $\sigma_{\ln M}\simeq 0.04$--0.06, depending on the
  physics included in the simulation, and with a mild dependence on
  redshift. As for the mass--temperature relation, it is the one with
  the largest scatter, with $\sigma_{\ln M}\magcir 0.1$ at $z=0$,
  increasing to $\magcir 0.15$ at $z=1$. The intrinsic scatter in the
  $M$--\yx\ relation is slightly larger than in the $M$--\mg\
  relation, with $\sigma_{\ln M}\simeq 0.06$ at $z=0$ and 0.08 at
  $z=1$.  These results confirm that both \yx\ and \mg\ mass proxies
  are well suited for cosmological applications of future large X--ray
  surveys. As a word of caution, we point out that the analysis
  presented in this paper does not include observational effects
  expected when measuring temperature from fitting X--ray spectra and
  gas mass from X--ray surface brightness profiles. A detailed
  assessment of such effects will be the subject of a forthcoming
  paper.
\end{abstract}



\section{INTRODUCTION}   
\label{sec:intro}
Measuring the mass of galaxy clusters represents the most important
aspect for the use of such structures as tracers of cosmic evolution.
The cosmological parameters depend, indeed, on the evolution of
cluster's mass function and on large--scale clustering properties
\citep[e.g.,][]{2001ApJ...561...13B,2002MNRAS.335..807S,Voit2005AdSpR..36..701V,Vikhlinin2009ApJ...692.1033V,Mantz2010MNRAS.tmp.1030M}.
Restricting the discussion to X--ray observations, to estimate cluster
masses we need to assume that the intra--cluster medium (ICM) sits in
hydrostatic equilibrium within the cluster potential well, generally
assumed to have spherical simmetry.  Further, we need to obtain gas
density and temperature profiles from high--quality X--ray data.
Limits on this approach are represented not only by the strong
assumptions made but also by the difficulty of obtaining data of
sufficient quality. This is especially true for distant clusters, at
$z\magcir 0.5$, for which observations of the needed sensitivity
would be prohibitively expensive in terms of required exposure time.

To overcome this problem one generally resorts to the observational
determinations of mass proxies, i.e. of suitably observable quantities
easier to measure and related to total cluster mass through the
so-called ``scaling relations''. Ideally, a robust mass proxy should be
characterized by a low intrinsic scatter with cluster mass, implying
that it neeeds to be relatively insensitive to the cluster dynamical
state. Different examples of X--ray mass proxies have been used over
the last decade to derive cosmological constraints from X--ray
clusters surveys. The easiest-to-measure mass proxy is the total
X--ray luminosity, $L_{\rm X}$. While this quantity is known to correlate
with cluster mass \citep[e.g.,][]{RE02.1,Pratt2009A&A...498..361P},
the $L_{\rm X}$--$M$ relation has a fairly large scatter, of about 40 per
cent, due to the sensitivity of X--ray luminosity on the details of
the gas distribution in central regions and, therefore, on the
dynamical state of the cluster. Other commonly adopted choices of mass
proxies are the X--ray temperature $T_{\rm X}$
\citep[e.g.][]{1991ApJ...372..410H,Markevitch1998ApJ...504...27M,
  2002A&A...383..773I,2003MNRAS.342..163P} and the cluster gas mass
\mg \citep[e.g.][]{2003ApJ...590...15V,2004ApJ...601..610V}. In order to 
estimate these quantities detailed X--ray brightess profiles must be 
obtained, requiring longer exposure times than those required by $L_{\rm X}$ 
only.
However, they are expected to be more robustly linked to total collapsed mass.

In this context cosmological hydrodynamical simulations are playing
an increasingly relevant role in calibrating the mass proxies, in
understanding the systematics on cluster mass measurements, and in
defining new mass proxies. Simulations provide the most advanced
theoretical tool to capture the complexity of the hierarchical
build--up of galaxy clusters and of the gas--dynamical processes
\citep[e.g.][for a recent review on cluster
  simulations]{Borgani2009arXiv0906.4370B}. A typical example is
provided by the role that simulations played in the last years to
calibrate the expected level of violation of hydrostatic equilibrium
induced by the presence of non--thermalised gas motions
\citep[e.g.,][]{2004MNRAS.351..237R,Nagai2007ApJ...655...98N,Jeltema2008ApJ...681..167J,
  Ameglio2009MNRAS.394..479A,Piffaretti2010arXiv1007.1916P,Lau2009ApJ...705.1129L}.
Another example is given by the application of simulations in the
study of the accuracy of X--ray temperature and of gas mass as tracers
of the total cluster mass
\citep[e.g.,][]{Evrard1996ApJ...469..494E,Bartelmann1996MNRAS.283..431B,
  Muanwong2006ApJ...649..640M,Ettori2004A&A...417...13E,Nagai2007ApJ...668....1N}.
With the increasing numerical resolution and degree of realism of the
physical processes included, these calibrations are becoming
trustworthy as convincingly shown by different works. A good match
between simulations and observations was shown for temperature
profiles of clusters \cite[e.g.,][]{Loken2002ApJ...579..571L,
  Borgani2004MNRAS.348.1078B,Kay2007MNRAS.377..317K,
  Nagai2007ApJ...668....1N,Leccardi2008A&A...486..359L}, gas density
profiles \cite[e.g.,][]{Roncarelli2006MNRAS.373.1339R,
  Croston2008A&A...487..431C} and pressure density
\cite[e.g.,][]{Arnaud2009arXiv0910.1234A}.

These comparison are mostly restricted to the region within $R_{500}$ 
\footnote{We define $M_\Delta$ as the mass contained within the radius
  $R_\Delta$ encompassing an overdensity of $\Delta$ times the cosmic
  critical density, $\rho_c(z)=3H(z)^2/8\pi G$.}, after excluding the
central regions affected by the complex physical processes
regulating the ``cool core'' structure of the ICM.

As for the cluster mass proxies simulations predict that the
$T_{\rm X}$--$M$ relation has an intrinsic scatter which is quite
sensitive to the presence of substructures in the ICM and to the
cluster dynamical state
\citep[e.g.,][]{OHara2006ApJ...639...64O,2009ApJ...699..315Y}.  The
gas mass proxy is instead expected to be less sensitive to the cluster
dynamical state, thus implying a smaller scatter in its scaling
relation with total mass. Moreover it has the advantage that it can be
essentially measured from X--ray imaging alone.  There is however a
caveat, since the dependence of gas mass fraction on cluster mass,
cluster-centric distance and on redshift are still not yet completely
understood \citep[e.g.,][]{Ettori2006MNRAS.365.1021E}.

\cite{Kravtsov2006ApJ...650..128K} introduced the X--ray equivalent of
the integrated SZ flux, the $Y_{\rm X}$ parameter, defined as the product of
the gas mass with the cluster temperature. From a set of sixteen
simulated galaxy clusters these authors found it to be a low--scatter
(5-8 per cent) mass proxy, due to the anticorrelation of the residuals
in the scaling relation of temperature and gas mass with total
mass. Furthermore, the scaling of $Y_{\rm X}$ with total mass was found to be
in close agreement with the prediction of the simple self--similar
model, based on the assumption that gas follows dark matter and
gravitational effects only determine the thermal content of the ICM
\citep[e.g.,][]{Kaiser1986MNRAS.222..323K}.  The $Y_{\rm X}$--$M$
relation is relatively insensitive to cluster mergers, a result that
was independently verified by \cite{Poole2007MNRAS.380..437P} and by
Rasia et al. (2010) with simulations of cluster mergers. A number of
recent observational works estimated $Y_X$ and its correlation with
X--ray luminosity.  Data from both XMM-Newton
\cite[][]{Pratt2009A&A...498..361P} and Chandra telescope
\cite[][]{Vikhlinin2009ApJ...692.1033V,Maughan2007ApJ...668..772M}
showed that the observed $Y_{\rm X}$--$M$ relation has slope and
redshift evolution in agreement with predictions of the self--similar
model.  With their sample of both nearby and distant clusters out to
$z\sim 1$, they confirm that gravitational processes are indeed
responsible for the bulk of the total thermal content of galaxy
clusters.

A more recent analysis of scaling relations for simulated ga\-la\-xy
clusters presented by \cite{Stanek2010} did not confirm the
anti--correlation of the residuals of $M--M_{\rm gas}$ and
$M-T_{\rm mw}$. They found that the lowest-scatter mass proxy is
$M_{\rm gas}$.  A similar result was also found by
\cite{Okabe2010arXiv1007.3816O} who computed the scaling relation
between X--ray mass proxies and the total cluster mass obtained from
lensing observations of 12 clusters at $z\sim 0.2$.  From their
analysis the conclusion is that $M_{\rm gas}$ has the lowest intrinsic
scatter of $\sim 10\%$, with respect to the $\sim 20\%$ of the $Y_{\rm X}$
proxy.

At present, $T_{\rm X}$, $M_{\rm gas}$ and $Y_{\rm X}$ 
are considered the most robust
indicators of cluster mass and, in fact, the most recent cosmological
applications of X--ray cluster surveys have been based on the use of
such mass proxies
\citep[e.g.,][]{Henry2009ApJ...691.1307H,Vikhlinin2009ApJ...692.1060V,Mantz2010MNRAS.tmp.1030M}.

Ideally, rather than relying on simulations, it would be preferrable
to calibrate mass proxies directly from observational data. An
observational calibration, while possible in principle, requires
highly precise measurements of cluster mass in a way which is
completely independent of X-ray observations, e.g. from gravitational
lensing or from the study of the cluster internal dynamics as traced
by member galaxies. Indeed, any determination of the scatter between
any of the above three X-ray mass proxies and cluster mass based on
the application of hydrostatic equilibrium would be just inconclusive,
owing to the fact that the relation is between two highly correlated
quantities. On the other hand, precise measurements of lensing masses
are now available only for a relatively small number of objects
\citep[e.g.,][]{Mahdavi2008MNRAS.384.1567M,Zhang2010ApJ...711.1033Z,
Okabe2010arXiv1007.3816O}. Although
progresses in this direction are expected in the future, simulations
represent today a valid alternative to calibrate mass proxies and to
understand their robustness.

A careful study of mass proxies through simulations requires a large
enough sample of clusters simulated both with adequate resolution and
by including a range of physical processes. The most direct way to
achieve large statistics of clusters would be to carry out simulations
of large cosmological boxes
\citep[e.g.][]{Borgani2004MNRAS.348.1078B,Gottloeber2006astro.ph..8289G,
Kay2007MNRAS.377..317K,Burns2008ApJ...675.1125B,Hartley2008MNRAS.386.2015H}. However,
these simulations have to severely compromise between box size,
achievable resolution and details with which different physical
processes (e.g., star formation and feedback from different sources)
can be studied. As an alternative, one could simulate at high
resolution only specific regions surrounding clusters, which are
previously identified from low-resolution simulations of large
cosmological volumes. The advantage of this approch is that it can
provide a more realistic numerical description, in terms of resolution
reached and of physical processes included, only within the
``zoomed-in'' Lagrangian regions surrounding clusters. As a matter of
fact, this resimulation procedure is non-trivial to implement, thus
limiting the statistics of resimulated clusters so-far presented in
the literature. For instance, \cite{Dolag2009MNRAS.399..497D}
presented a set of $18$ such clusters, of which however only $4$ have
$M_{200}>10^{15} \msun$ \citep[see also][]{Fabjan2010MNRAS.401.1670F}.
\cite{Lau2009ApJ...705.1129L} and \cite{Puchwein2008ApJ...687L..53P}
presented sets of resimulated clusters of comparable size, again with
only a couple of them being as massive as $\sim 10^{15} \msun$.

In this paper we use two sets of simulated clusters, whose combination
provides us with both a large statistics of massive objects and a
range of physical processes included over which to test the stability
of the scaling relations. The first set of clusters (Set 1 hereafter)
contains about 140 clusters, 30 of which have masses larger than
$10^{15}\msun$. For all of them, we carried out simulations with both
non--radiative physics and by including cooling, star formation and
galactic winds powered by Type-II supernova (Sn-II) explosions. The
second set of simulations (Set~2 hereafter) includes 18 clusters, with
4 of them having masses larger than $10^{15}\msun$. We refer to
\cite{Dolag2009MNRAS.399..497D} for a complete description of the
second set of simulations. This smaller set of simulated clusters
complements the first set since it has been simulated by changing in
seven different ways the description of physical processes
(e.g. artificial viscosity, cooling, star formation, feedback
efficiency from SN and from gas accretion onto supermassive black
holes). The complementary analysis of these two cluster sets allows
us to address in detail the stability and robustness in the
calibration of X--ray mass proxies with simulations, by combining
statistics of massive halos, numerical resolution and range of
physical processes included.

In this first paper we will carry out an analysis of scaling relation
of mass proxies against mass by neglecting all observational
effects. For this reason ICM temperature will be mass--weighted (where
not listed differently), and all quantities will be estimated in a
three--dimensional analysis without including projection effects. In
this way, we will assess the intrinsic performances of $T_{\rm X}$,
$M_{\rm gas}$ and $Y_{\rm X}$ as tracers of the true mass, while we defer to a
forthcoming paper (Rasia et al., in prep.) a detailed analysis of the
impact of observational biases.

This paper is organized as follows. We present in Section
\ref{sec:simul} the simulations analysed in this paper. After a
description of the \gadget simulation code
\citep{Springel2005MNRAS.364.1105S} and of the physical processes
included in the simulations, we describe the initial conditions and
the general characteristics of the resulting samples of simulated
clusters. In Section \ref{sec:res} we present the results of our
analysis. After introducing the definitions of mass proxies, we
present results at $z=0$, along with an assessment of the dependence
of the scaling relations on the physical processes included in the
simulations. We finally discuss the evolution of scaling relations and
of their intrinsic scatter with redshift. We discuss our results and
summarize the main conclusions of our analysis in Section
\ref{sec:conc}.


\section{Simulations} 
\label{sec:simul}
\subsection{The simulation code and adopted physical models}

In this Section we introduce the simulation code 
and all the different physical processes considered in re-simulating
our objects. Simulations have been carried out using the 
Tree--PM SPH \gadget \citep{Springel2005MNRAS.364.1105S} code. 
Oldest \gadget~2 version was used for Set 2 simulations, already 
presented by \cite{Dolag2009MNRAS.399..497D}, while 
Set 1 was completed using the newest version 3. 
The difference between the two versions resides mainly
in the different algorithm adopted for domain decomposition. Both
versions of the code use segments of the space-filling Peano--Hilbert
curve to decide the particles to be assigned to different
processors. Unlike \gadget~2, the newest \gadget~3 version allows
each processor to be assigned also disjoined segments of the
Peano--Hilbert curve. This turns into a substantial improvement of the
work--load balance assigned to the different processors, especially for
simulations, like those presented here, in which the computation cost
is largely concentrated within a quite small fraction of the physical
volume of the computational domain. 

We add here below a brief description for each of the physical models 
adopted within our simulations. 

\begin{description}

\item{\nrsv\ (non-radiative, standard viscosity):} 

non-radiative runs, using the standard reference scheme
  for artificial viscosity implemented in \gadget\. This prescription
  is based on the formulation of artificial viscosity originally
  presented by \cite{Monaghan1997JCoPh.136..298M}, also including a
  viscosity limiter as proposed by \cite{Balsara1995JCoPh.121..357B}
  and \cite{Steinmetz1996MNRAS.278.1005S}.

\item{\nrrv\ (non radiative, reduced viscosity):}

 non-radiative runs, using reduced time--dependent
  viscosity, as originally proposed by
  \cite{Morris1997JCoPh.136...41M} and implemented in \gadget by
  \cite{Dolag2005MNRAS.364..753D}. In this scheme, artificial
  viscosity decays away from shock regions, thus allowing a higher
  degree of turbulence in the velocity field to develop.

\item{\csf\ (cooling star formation and feedback):}
 
  runs including radiative cooling for a zero--metallicity
  plasma, including heating/cooling from a spatially uniform and
  evolving UV background \citep{Haardt1996ApJ...461...20H}. Gas
  particles above a given threshold density are treated as multiphase,
  so as to provide a sub--resolution description of the inter--stellar
  medium, according to the model described by
  \cite{Springel2003MNRAS.339..289S}. Within each multiphase gas
  particle, a cold and a hot-phase cohexist in pressure equilibrium,
  with the cold phase providing the reservoir of star
  formation. Kinetic feedback is implemented by giving some SNII energy 
  to gas particles in the form of kinetic energy, thus mimicking galactic
  ejecta powered by SN explosions. In these runs, galactic winds have
  a mass upload proportional to the local star-formation rate. We use
  $v_w=340\vel$ for the wind velocity, which corresponds to assuming
  half of energy released by SN-II being converted to kinetic energy
  for a Salpeter IMF \citep{Salpeter1955ApJ...121..161S}.

\item{\csfc\ (csf and thermal conduction):} 

the same as the \csf\ runs, but also including the
  effect of thermal conduction, as described by
  \cite{Jubelgas2004MNRAS.351..423J} and
  \cite{Dolag2004ApJ...606L..97D}. The conduction efficiency is 
  assumed to have a value of $1/3$ of the Spitzer conductivity.

\item{\w\ (csf, metals and galactic winds):} 

radiative runs with star formation described through the
  same multiphase model of the \csf\ runs. In addition, \w\
  runs also include a description of metal production from chemical
  enrichment contributed by SN-II, SN-Ia and AGB stars, as described
  by \cite{Tornatore2007MNRAS.382.1050T}. Stars of different mass,
  distributed according to a Salpeter IMF \citep{Salpeter1955ApJ...121..161S}, 
  release metals over the
  time-scale determined by the corresponding mass-dependent life-times
  (taken from \citealt{Padovani1993ApJ...416...26Pb}). The metalli\-city
  dependence of radiative cooling is included by using the cooling
  tables by \cite{Sutherland1993ApJS...88..253S}. In these runs, the
  velocity of galactic ejecta is assumed to be $v_w=500\vel$, thus
  corresponding to a more efficient SN feedback that in the \csf\
  runs.

\item{\nw\ (csf, metals and no winds):} 

the same as \w\ but with no winds, i.e. by
  ex\-clu\-ding the effect of kinetic feedback from galactic winds.

\item{\agn\ (csf, metals and AGN):} 

the same as the \w\ runs, but replacing SN feedback in
  the form of galactic winds, with the effect of AGN feedback
  triggered by gas accretion onto supermassive black holes
  (BH). 
  Details about this feedback model are discussed in 
  \cite{Fabjan2010MNRAS.401.1670F},
  while we summarize here only the main points. 
  The scheme is a slight modification of the model introduced by
  \cite{Springel2005MNRAS.361..776S}, where BH are represented
  by sink particles initially seeded in resolved DM halos. These
  particles increase their mass by gas accretion and merging with
  other BH particles. Eddington-limited Bondi accretion produces a
  radiated energy which correspond to a fraction $\epsilon_r=0.1$ of
  the rest-mass energy of the accreted gas. A fraction $\epsilon_f$ of
  this radiated energy is thermally coupled to the surrounding gas. We
  use $\epsilon_f=0.05$ for this feedback efficiency, which increases
  to $\epsilon_f=0.2$ whenever accretion enters in the quiescent
  ``radio'' mode and takes place at a rate smaller than one-hundred of
  the Eddington limit
  \citep[e.g.][]{Sijacki2006MNRAS.366..397S,Fabjan2010MNRAS.401.1670F}.
 \end{description}

 We remark here that galaxy clusters in Set 2 were simulated with all 
 the presented baryon physics models, while simulations of Set 1 
 clusters were performed with \nrsv\ and \w\ physics only.

 The diversity and heterogeneity of the description of physical
 processes determining the evolution of the ICM is not meant to
 provide a systematic study of the parameter space describing 
 the ICM physics. However, quantifying the
 variation of the scaling relations between mass proxies and true
 cluster mass for each of such models will allow us to assess the
 robustness of mass proxies against the uncertainties in the
 description of the ICM physics.

\subsection{Initial conditions and simulation sets} 

Our analysis is based on two sets of simulations of galaxy
clusters. The two sets, while having similar mass and force
re\-so\-lu\-tion, are quite different in a number of aspects and allow us to
perform complementary tests on the robustness of mass proxies.
The large statistics of Set 1 clusters allows us
to ca\-li\-bra\-te with precision the slope and scatter of the scaling relations 
along with their evolution.
While the variety of different physical processes of Set 2 
are used to assess the robustness of mass proxies against the
uncertainties in the description of the ICM physics.

We provide in the following a more technical description of these two sets of
simulations.

\subsubsection{Set 1} is based on simulations of 29 Lagrangian regions
extracted around as many clusters identified within a low--re\-so\-lution
N-body cosmological simulations. We provide here below a short
description of these initial conditions, while a more detailed
presentation will be provided in a forthcoming paper (Bo\-na\-fe\-de et 
al., in preparation). The parent DM simulation follows
$1024^3$ DM particles within of a box having a comoving side of
$1\, h^{-1}$Gpc. The cosmological model assumed is a flat $\Lambda$CDM
one, with $\Omega_m=0.24$ for the matter density parameter,
$\Omega_{\rm bar}=0.04$ for the contribution of baryons,
$H_0=72\,{\rm km\,s^{-1}Mpc^{-1}}$ for the present-day Hubble
constant, $n_s=0.96$ for the primordial spectral index and
$\sigma_8=0.8$ for the normalization of the power spectrum in terms of
the r.m.s. fluctuation level at $z=0$ within a top-hat sphere of
$8\hm$ radius. With the above choice of parameters, this cosmological
model is consistent with the CMB WMAP-7 constraints
\citep{Komatsu2010arXiv1001.4538K}. The selected Lagrangian regions
have been chosen so that 24 of them are centered around the 24 most
massive clusters found in the cosmological volume, all having
virial\footnote{Here we define the virial mass $M_{\rm vir}$, as the
  mass contained within the radius encompassing the overdensity of
  virialization, as predicted by the spherical collapse model
  \citep[e.g.,][]{Eke1996MNRAS.282..263E}.} mass $M_{\rm vir}\magcir
10^{15}\msun$. Further five regions have been chosen around clusters
in the mass range $M_{\rm vir}\simeq (1-7)\times 10^{14}\msun$.

\begin{figure}
\hspace{-1.truecm}
\psfig{figure=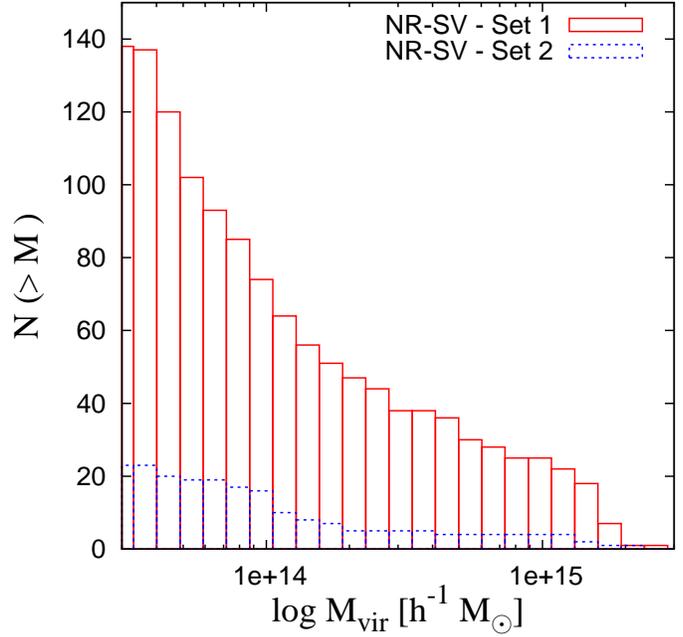,width=10.cm,angle=-90}
\caption{The cumulative distributions of virial masses for clusters in
  Set 1 (red solid histogram) and in Set 2 (blue dashed histogram). We
  only count ``clean'' clusters, for which no contaminant
  low--resolution DM particles are found within their corresponding
  virial radii.}
\label{fig:nclus}
\end{figure}

Within each Lagrangian region we increased mass resolution and added
the relevant high-frequency modes of the power spectrum, following the
Zoomed Initial Condition (ZIC) technique
\citep{Tormen1997MNRAS.286..865T}. Initial conditions for hydrodynamic
simulations have been generated by splitting each particle within the
high--resolution region into a DM and a gas particle, having mass ratio
such to reproduce the cosmic baryon fraction. Outside high--resolution
regions resolution is progressively degraded, so as to save
computational time, while preserving a correct description of the
large--scale tidal field. We generated initial conditions at
different, progressively increasing, resolution. For the hydrodynamic
simulations presented here 
the mass of each DM particle in the
high-resolution region is $m_{\rm DM}\simeq 8.47\times 10^8\msun$, with
$m_{\rm gas}\simeq 1.53\times 10^8\msun$ for the initial mass of gas
particles. Using an iterative procedure, we have shaped each
high--resolution Lagrangian region in such a way that no low--resolution
particle contaminates the central ``zoomed in'' halo at $z=0$ at least
out to 5 virial radii of the central cluster. This implies that each
region is sufficiently large to contain more than one interesting
cluster with no ``contaminants'' out to at least one virial radius.
In total, we find $\simeq 140$ clusters with $M_{500}\ge
5\times 10^{13}\msun$, out of which about 40 have $M_{\rm vir}>
5\times 10^{14}\msun$ and 30 have $M_{\rm vir}> 10^{15}\msun$. We show
in Fig.~\ref{fig:nclus} the cumulative mass distribution for this
set of simulated clusters. 

Simulations have been carried out using a Plummer--equivalent softening
length for the computation of the gravitational force in the
high--resolution region which is fixed to $\epsilon=5\,h^{-1}$kpc
in physical units at redshift $z<2$, while being kept fixed in
comoving units at higher redshift. As for the computation of
hydrodynamic forces, we assume the SPH smoothing length to reach a
minimum allowed value of $0.5\epsilon$.

Clusters from this set have been simulated using the \nrsv\ and \w\
models. The \nrsv\ set is the only one simulated with \gadget~3 code.

\subsubsection{Set 2} 
is based on 9 Lagrangian
regions selected from a parent DM only cosmological box with a size of
$479 \, h^{-1}$Mpc \cite{Yoshida2001MNRAS.328..669Y}. The cosmological
model assumed is $\Lambda$CDM with $\Omega_M=0.3$ for the matter
density, $h=0.7$ for the Hubble parameter, $\Omega_{\rm bar}=0.039$
for the baryon density parameter and $\sigma_8$=0.9 for the
normalization of the power spectrum. We refer to
\cite{Dolag2009MNRAS.399..497D} for a more detailed description of
this set of clusters. Briefly, this set includes overall 18 clusters
with $M_{\rm vir}> 5\times 10^{13}\msun$, out of which only 4 have
$M_{\rm vir}> 10^{15}\msun$. The mass distribution of these clusters
is shown in Fig.~\ref{fig:nclus}. 

Mass resolution is quite close to that of Set 1 
($m_{\rm DM}\simeq 1.9\times 10^8\msun$ and  
$m_{\rm gas}\simeq 2.8\times 10^7\msun$ for DM and gas particles respectively),
with also the same
choices for the size of the gravitational softening scale and for the
minimum allowed value of the SPH smoothing length.  Clusters from this
set have been simulated with \gadget~2 using all the seven above
described different models for the description of the baryon phy\-sics.

\begin{figure*}
\hbox{
\hspace{-0.2truecm}
\psfig{figure=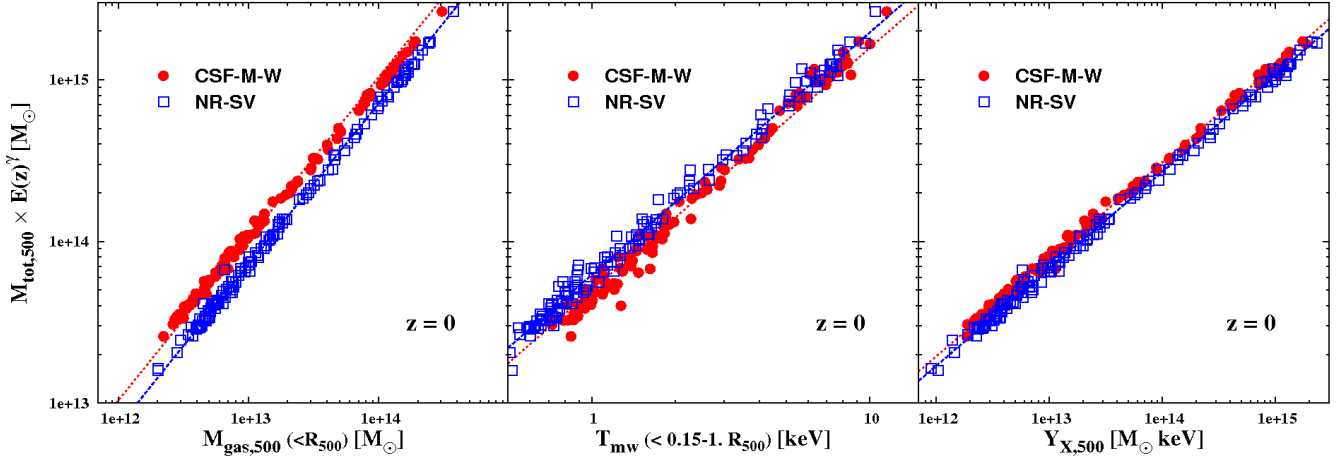,width=18.0cm,angle=0}
}
\caption{Scaling relations at $z=0$ between the total mass computed
  within $R_{500}$ and three mass proxies: \mg, \tmw\ and \yx. 
  The evolution of the relation is scaled accordingly, with
  $\gamma = 0, 1$ and $2/5$, respectively. On each figure we plot the
  results for non-radiative (blue open squares) and radiative (red filled
  circles) runs for clusters in Set 1 with $M_{\rm vir}> 5\times
  10^{13}\msun$. The short dashed (red) and dashed (blue) lines are
  the best fit relations of respectively radiative and non--radiative
  cluster simulations with slopes fixed to the self--similar value.
  Left panel: scaling with \mg; central panel: scaling with
  mass--weighted temperature, \tmw; right panel: scaling with \yx. The
  values of \tmw\ appearing in the central panel and entering in the
  computation of \yx\ in the right panel are computed within the
  radial range (0.15--1)$R_{500}$.}
\label{fig:SRred0}
\end{figure*}

\begin{figure*}
\hbox{
\hspace{-0.2truecm}
\psfig{figure=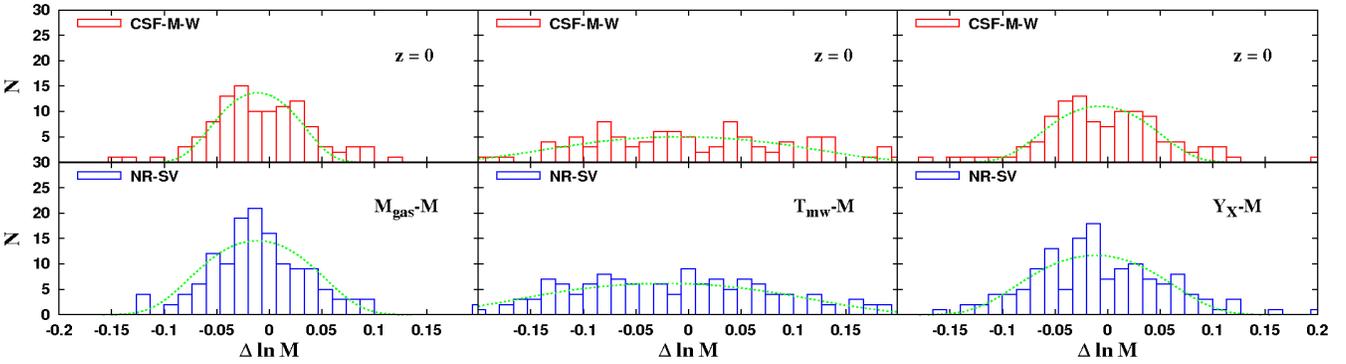,width=18.0cm}
}
\caption{Distributions of the residuals in log--mass in the scaling
  relations at $z=0$ between mass proxies and $M_{\rm tot,500}$, for the
  non--radiative (upper red histograms) and radiative (lower blue
  histograms) runs of simlated clusters from Set 1. Residuals are
  computed with respect to the relation that fits at best the
  simulated clusters data. Left panel: scaling with \mg; central
  panel: scaling with mass--weighted temperature, \tmw; right panel:
  scaling with \yx. Dotted curves in green show the Gaussian 
  distribution having the same variance of
  the distribution of residuals in mass (as reported in Tables
  \protect\ref{tab:YX}, \protect\ref{tab:Mgas} and
  \protect\ref{tab:Tmw}).}
\label{fig:scat_z0}
\end{figure*}

\section{Results and discussion}
\label{sec:res}
\subsection{Mass proxies}
The mass proxies that we consider in our analysis are the
mass-weighted temperature $T_{mw}$, the gas mass $M_{gas}$ and the
product of gas mass and temperature, $Y_X=M_{gas} \times T_{mw}$. All
these quantities are computed within the fiducial radius $R_{500}$,
which typically corresponds to the outermost radius where scaling
relations from observational data analysis are provided
\citep[e.g.,][]{Vikhlinin2009ApJ...692.1033V,Sun2009ApJ...693.1142S,
  Pratt2009A&A...498..361P,Maughan2007ApJ...668..772M}.  With the
purpose of reproducing the procedure of observational analyses we
exclude the central regions, $R<0.15R_{500}$, in the computation of
the temperature. The reason for this choice in the analysis of real
clusters is that temperature profiles show a rather high degree of
diversity in such central regions depending on their degree of
``cool--coreness''
\citep[e.g.,][]{Leccardi2008A&A...486..359L,Vikhlinin2006ApJ...640..691V,
  Pratt2007A&A...461...71P}.  Excluding such regions suppresses the
scatter in the scaling relation involving temperature
\citep[e.g.,][and references therein]{Pratt2009A&A...498..361P}.
Furthermore, it places us in the regime where simulations are in
closer agreement with observations
\citep[e.g.,][]{Borgani2004MNRAS.348.1078B}.  The choice of the radial
interval where to compute mass proxies represents our only attempt to
introduce ``observational effects'' in our analysis. Indeed, in this
paper (Paper I) we prefer to analyse in detail the intrinsic
performances of the three above mass proxies, i.e. neglecting any
possible observational bias which will be addressed in 
Paper II (Rasia et al., in prep.). We want to stress out that in this
analysis i) we estimate all quantities in three-dimensions (i.e.,
neglecting projection effects); ii) unless otherwise stated, we use
the mass--weighted definition of temperature; iii) we do not remove
gas clumps at relatively low temperature when we compute both
temperature and gas mass.

The simplest description of scaling relations between galaxy cluster
mass and X--ray observables is that provided by the self--similar
model originally proposed by
\citep{Kaiser1986MNRAS.222..323K}. According to this model, the
thermodynamical properties of the ICM are determined only by the
action of gravity, under the assumption of virial equilibrium and
spherical simmetry \citep[e.g.,][ for a
review]{Voit2005RvMP...77..207V}. Within this model cluster total mass
is the only parameter that determines all the thermodynamical
properties of the hot intra--cluster gas. Accordingly, the cluster gas
mass within the radius $R_\Delta$ is proportional to the total mass
computed within the same radius,
\be
M_{tot,\Delta} = C_{M_g} M_{gas,\Delta} \,,
\ee
with constant of proportionality $C_{M_g}$ being independent of redshift.
The total mass is related to the temperature according to 
\be 
M_{tot,\Delta} = E(z)^{-1} C_T T_\Delta^{3/2} \, ,
\label{eq:SR} 
\ee
where $E(z)=H(z)/H_0$ defines the evolution of the Hubble parameter. 
Therefore, as defined in the paper by \cite{Kravtsov2006ApJ...650..128K} 
the self--similar scaling of $Y_X$ with cluster mass is
given by 
\be 
M_{tot,\Delta} = C_{Y_X} E(z)^{-2/5} Y_{X,\Delta}^{3/5}\, .  
\ee 
In the above relations involving ICM temperature the mass--averaged
\tmw\ should be used, since it is more directly related to the total
ICM thermal content. As such, this is the quantity relevant for the
predictions of the self--similar model, rather than the spectroscopic
temperature which is more affected by the complexity of the ICM
thermal structure.

In the following we analyse the scaling relations for simulated
clusters with masses $M_{vir} > 5 \times 10^{13} \msun$, at five
different redshifts, $z=0, 0.25, 0.5, 0.8$ and $1$.  To fit the
scaling relation between total mass, $M_{tot,500}$, and a given
observable mass proxy $X_{500}$, we use a power law
\be
M_{tot,500} = C_X \left( \frac{X_{500}}{X_0} \right)^{\alpha_X} \, .
\label{eq:M_X}
\ee
Following the work by \cite{Kravtsov2006ApJ...650..128K} the
normalization factor $X_0$ was fixed to $2 \times 10^{13} M_{\odot}$,
$3$ keV and $4 \times 10^{13}$ keV $M_{\odot}$ for $M_{gas}$, $T_{mw}$
and Y$_X$, respectively. We also perform the fits by fixing the slope
$\alpha_X$ to the self--similar value of $\alpha_{SS} = 1.0, 1.5$ and
$0.6$ for \mmg, \mt\ and \myx, respectively.

The values of the slope $\alpha_X$ and normalization $C_X$ are
obtained by performing a best--fitting on the log--log relation by
using the least-squares Marquardt-Levenberg algorithm
\citep{1992nrca.book.....P}. This algorithm calculates at each
iteration the sum of the squared differences with a new set of
parameter values. The Marquardt-Levenberg algorithm selects the
parameter values for the next iteration. The process continues until a
preset criterium is met, either the fit has converged (the relative
change in the residuals is less than $10^{-6}$) or it reaches a preset
iteration count limit. The intrinsic scatter in natural logarithm of
the total mass, $\ln M_{\rm tot}$, around each best--fit scaling relation
is computed as the quadratic difference between the single cluster
data and the relation fitted on the whole cluster set.


\subsection{Scaling relations at $z=0$}

In this section we will discuss the behavior of \mmg, 
\mt\ and \myx\ relations for the radiative (\w) and
non--radiative simulations (\nrsv) at $z=0$ for the simulated clusters
of Set 1.

In Fig.~\ref{fig:SRred0} we show the relations between the total
cluster mass $M_{\rm 500}$ and \mg\ (left panel), \tmw\ (middle panel),
$Y_X$ (right panel). Besides the results of the analysis for radiative
(\w\ in red circles) and non--radiative (\nrsv\ in blue squares)
simulations, we plot the best--fit relations obtained when the slope
is fixed to the corresponding self-similar values, shown with the red
long-dashed and the blue short-dashed lines for the \w\ and \nrsv\
simulations, respectively.

Galaxy clusters in non--radiative simulations show a behaviour close
to the self--similar one for the three studied relations. Indeed, when
we consider the slope as free parameter the new best fits (not shown)
over all \nrsv\ clusters return values consistent with $\alpha_{SS}$,
namely, $\alpha=0.981\pm0.004$ for \mg, $\alpha=1.517\pm0.012$ for
\tmw\ and $\alpha=0.597\pm0.03$ \yx, respectively. Results of the
fitting relations are repoted in Tables \ref{tab:YX}, \ref{tab:Mgas},
and \ref{tab:Tmw} for the \yx, \mg, and \tmw, respectively.  The
agreement with the self--similar scaling for \tmw\ was found also by
\cite{Stanek2010} for clusters extracted from the non--radiative
version of the Millenium Gas Simulations (MGS).

When radiative cooling and star formation are included in \w\
simulations, the average quantities change thus modifying slope and
normalization of the scaling relations. The \w\ clusters (red
circles in Fig.~\ref{fig:SRred0}) deviate from the self-similar
scaling relation for both \mg\ and \tmw. As for the \mmg\
scaling, the best fit relation has a shallower slope. This is due to
the fact that for radiative simulations the conversion of baryons into
stars is relatively more efficient in lower mass clusters, thus
corresponding to a lower mass fraction of gas left in the hot diffuse
phase
\citep[e.g.][]{Borgani2004MNRAS.348.1078B,Kravtsov2005ApJ...625..588K,Fabjan2010MNRAS.401.1670F}.
As for the \mt\ relation, the results for the radiative
runs are characterized by a slope steeper than the self--similar one,
with $\alpha=1.615\pm0.016$. In this case, the effect of including
radiative physics is that of increasing the temperature of the ICM by
a larger amount in less massive systems. Indeed, the more efficient
cooling in lower--mass systems causes a relatively stronger adiabatic
compression of gas in central core regions, as a consequence of the
lack of pressure support. As a result, temperature profiles in
radiative simulations of clusters are relatively steeper in less
massive systems \citep[see also][]{Borgani2004MNRAS.348.1078B}. 

\begin{table*}
\begin{center}
\begin{tabular}{lccccc}
\hline
     & $z = 0$ & $z = 0.25$ & $z = 0.50$ & $z = 0.80$ & $z = 1$ \\ 
\hline
\hline
\\
\multicolumn{6}{c}{\nrsv\ best--fitting parameters} 
\\
$\alpha_{Y_X}$  	 &  0.597 $\pm$  0.003 	 &  0.599 $\pm$  0.003 	 &  0.604 $\pm$  0.003 	 &  0.601 $\pm$  0.004 	 &  0.604 $\pm$  0.005 	 \\
$\log C_{Y_X}$ 	 &  14.190 $\pm$  0.002 	 & 14.188 $\pm$  0.002 	 & 14.189 $\pm$  0.002 	 & 14.177 $\pm$  0.003 	 & 14.185 $\pm$  0.004 	 \\
$\sigma_{\ln M}$ 	 &  0.064 	 	 &  0.067 	 	 &  0.067 	 	 &  0.070 	 	 &  0.073 	 	 \\
$\sigma_{\ln Y_X}$ 	 &  0.107 	 	 &  0.111 	 	 &  0.111 	 	 &  0.117 	 	 &  0.122 	 	 \\

\hline
\\
\multicolumn{6}{c}{\nrsv\ self--similar scaling ($\alpha_{Y_X}=0.6$)} 
\\
$\log C_{Y_X}$ 	 &  14.191 $\pm$  0.002 	 & 14.188 $\pm$  0.002 	 & 14.187 $\pm$  0.002 	 & 14.177 $\pm$  0.002 	 & 14.183 $\pm$  0.003 	 \\
$\sigma_{\ln M}$ 	 &  0.064 	 	 &  0.067 	 	 &  0.067 	 	 &  0.070 	 	 &  0.074 	 	 \\
$\sigma_{\ln Y_X}$ 	 &  0.107 	 	 &  0.111 	 	 &  0.112 	 	 &  0.117 	 	 &  0.123 	 	 \\

\hline
\hline
\\
\multicolumn{6}{c}{\w\ best--fitting parameters} 
\\
$\alpha_{Y_X}$	 &  0.591 $\pm$  0.003 	 &  0.590 $\pm$  0.002 	 &  0.596 $\pm$  0.003 	 &  0.591 $\pm$  0.003 	 &  0.596 $\pm$  0.005 	 \\
$\log C_{Y_X}$	 &  14.249 $\pm$  0.002 	 & 14.247 $\pm$  0.002 	 & 14.250 $\pm$  0.002 	 & 14.248 $\pm$  0.003 	 & 14.252 $\pm$  0.004 	 \\
 $\sigma_{\ln M}$ 	 &  0.050 	 	 &  0.051 	 	 &  0.054 	 	 &  0.061 	 	 &  0.076 	 	 \\
 $\sigma_{\ln Y_X}$ 	 &  0.084 	 	 &  0.086 	 	 &  0.091 	 	 &  0.104 	 	 &  0.128 	 	 \\

\hline
\\
\multicolumn{6}{c}{\w\ self--similar scaling ($\alpha_{Y_X}=0.6$)} 
\\
$\log C_{Y_X}$ 	 &  14.250 $\pm$  0.003 	 & 14.250 $\pm$  0.002 	 & 14.252 $\pm$  0.002 	 & 14.253 $\pm$  0.002 	 & 14.255 $\pm$  0.002 	 \\
$\sigma_{\ln M}$ 	 &  0.052 	 	 &  0.054 	 	 &  0.055 	 	 &  0.063 	 	 &  0.076 	 	 \\
$\sigma_{\ln Y_X}$ 	 &  0.087 	 	 &  0.089 	 	 &  0.091 	 	 &  0.104 	 	 &  0.127 	 	 \\

\hline
\end{tabular}
\end{center}
\caption{Fitting parameters for the \myx\ relation for simulated clusters of Set 1. Results are obtained through a log--log linear regression using eq.(\protect\ref{eq:M_X}), with $Y_{\rm X,0}=4\times 10^{13}{\rm keV}\,{\rm M}_\odot$. Reported errors on the fitting parameters correspond to 1$\sigma$ standard deviation in the linear fit. Also reported are the values of the r.m.s. scatter in $M_{\rm tot,500}$ and in $Y_{\rm X,500}$ around the fitting relations. Different columns refer to results at different redshifts. First two lines report results for non--radiative (\protect\nrsv) simulations, the other two instead the results for radiative (\protect\w) simulations. For each simulation we report first the best--fitting parameters and below the results when the slope of the scaling relation is fixed at the value predicted by the self-similar model.}

\label{tab:YX}
\end{table*}

\begin{table*}
\begin{center}
\begin{tabular}{lccccc}
\hline
      & $z = 0$ & $z = 0.25$ & $z = 0.50$ & $z = 0.80$ & $z = 1$ \\
\hline
\hline
\\
\multicolumn{6}{c}{\nrsv\ best--fitting parameters} \\ 
$\alpha_{M_{gas}}$   	 &  0.981 $\pm$  0.004 	 &  0.983 $\pm$  0.003 	 &  0.991 $\pm$  0.004 	 &  0.991 $\pm$  0.004 	 &  0.999 $\pm$  0.005 	 \\
$\log C_{M_{gas}}$ 	 &  14.154 $\pm$  0.002 	 & 14.142 $\pm$  0.002 	 & 14.136 $\pm$  0.002 	 & 14.130 $\pm$  0.002 	 & 14.127 $\pm$  0.002 	 \\
$\sigma_{\ln M}$  	 &  0.055 	 	 &  0.048 	 	 &  0.047 	 	 &  0.041 	 	 &  0.040 	 	 \\
$\sigma_{\ln M_{gas}}$  	 &  0.056 	 	 &  0.049 	 	 &  0.047 	 	 &  0.041 	 	 &  0.040 	 	 \\

\hline
\\
\multicolumn{6}{c}{\nrsv\ self--similar scaling ($\alpha_{Y_X}=1$)} \\
$\log C_{M_{gas}}$ 	 &  14.155 $\pm$  0.002 	 & 14.144 $\pm$  0.002 	 & 14.138 $\pm$  0.002 	 & 14.132 $\pm$  0.001 	 & 14.128 $\pm$  0.001 	 \\
$\sigma_{\ln M}$  	 &  0.061 	 	 &  0.052 	 	 &  0.048 	 	 &  0.042 	 	 &  0.040 	 	 \\
$\sigma_{\ln M_{gas}}$  	 &  0.061 	 	 &  0.052 	 	 &  0.048 	 	 &  0.042 	 	 &  0.040 	 	 \\

\hline
\hline
\\
\multicolumn{6}{c}{\w\ best--fitting parameters} \\
$\alpha_{M_{gas}}$   	 &  0.929 $\pm$  0.003 	 &  0.926 $\pm$  0.003 	 &  0.929 $\pm$  0.003 	 &  0.924 $\pm$  0.004 	 &  0.933 $\pm$  0.005 	 \\
$\log C_{M_{gas}}$ 	 &  14.310 $\pm$  0.002 	 & 14.298 $\pm$  0.002 	 & 14.291 $\pm$  0.002 	 & 14.285 $\pm$  0.002 	 & 14.288 $\pm$  0.003 	 \\
$\sigma_{\ln M}$  	 &  0.039 	 	 &  0.043 	 	 &  0.039 	 	 &  0.046 	 	 &  0.047 	 	 \\
$\sigma_{\ln M_{gas}}$  	 &  0.042 	 	 &  0.047 	 	 &  0.042 	 	 &  0.050 	 	 &  0.050 	 	 \\

\hline
\\
\multicolumn{6}{c}{\w\ self--similar scaling ($\alpha_{Y_X}=1$)} \\
$\log C_{M_{gas}}$ 	 &  14.317 $\pm$  0.004 	 & 14.320 $\pm$  0.003 	 & 14.320 $\pm$  0.002 	 & 14.320 $\pm$  0.002 	 & 14.322 $\pm$  0.002 	 \\
$\sigma_{\ln M}$  	 &  0.090 	 	 &  0.092 	 	 &  0.078 	 	 &  0.080 	 	 &  0.068 	 	 \\
$\sigma_{\ln M_{gas}}$  	 &  0.090 	 	 &  0.092 	 	 &  0.078 	 	 &  0.080 	 	 &  0.068 	 	 \\

\hline
\end{tabular}
\end{center}
\caption{The same as in Table \protect\ref{tab:YX}, but for the
  \mmg\ relation. In this case we use
  $M_{\rm gas,0}=2\times 10^{13} M_\odot$ for the normalization factor in
  eq.(\protect\ref{eq:M_X}).}
\label{tab:Mgas}
\end{table*}

\begin{table*}
\begin{center}
\begin{tabular}{lccccc}
\hline
  & $z = 0$ & $z = 0.25$ & $z = 0.50$ & $z = 0.80$ & $z = 1$ \\
\hline
\hline
\\
\multicolumn{6}{c}{\nrsv\ best--fitting parameters} \\
$\alpha_{T_{mw}}$   	 &  1.517 $\pm$  0.012 	 &  1.515 $\pm$  0.016 	 &  1.534 $\pm$  0.016 	 &  1.498 $\pm$  0.021 	 &  1.495 $\pm$  0.027 	 \\
$\log C_{T_{mw}}$	 &  14.513 $\pm$  0.006 	 & 14.524 $\pm$  0.007 	 & 14.535 $\pm$  0.007 	 & 14.507 $\pm$  0.009 	 & 14.530 $\pm$  0.010 	 \\
$\sigma_{\ln M}$  	 &  0.120 	 	 &  0.145 	 	 &  0.134 	 	 &  0.146 	 	 &  0.150 	 	 \\
$\sigma_{\ln T _{mw}}$  	 &  0.079 	 	 &  0.096 	 	 &  0.087 	 	 &  0.098 	 	 &  0.101 	 	 \\

\hline
\\
\multicolumn{6}{c}{\nrsv\ self--similar scaling ($\alpha_{Y_X}=1.5$)} \\
$\log C_{T_{mw}}$	 &  14.508 $\pm$  0.005 	 & 14.519 $\pm$  0.005 	 & 14.524 $\pm$  0.004 	 & 14.508 $\pm$  0.005 	 & 14.531 $\pm$  0.005 	 \\
$\sigma_{\ln M}$ 	 &  0.120 	 	 &  0.145 	 	 &  0.135 	 	 &  0.146 	 	 &  0.150 	 	 \\
$\sigma_{\ln T_{mw}}$ 	 &  0.080 	 	 &  0.097 	 	 &  0.090 	 	 &  0.098 	 	 &  0.100 	 	 \\

\hline
\hline
\\
\multicolumn{6}{c}{\w\ best--fitting parameters} \\
$\alpha_{T_{mw}}$   	 &  1.615 $\pm$  0.016 	 &  1.614 $\pm$  0.013 	 &  1.640 $\pm$  0.017 	 &  1.608 $\pm$  0.019 	 &  1.596 $\pm$  0.028 	 \\
$\log C_{T_{mw}}$ 	 &  14.427 $\pm$  0.006 	 & 14.436 $\pm$  0.005 	 & 14.450 $\pm$  0.006 	 & 14.445 $\pm$  0.006 	 & 14.444 $\pm$  0.009 	 \\
$\sigma_{\ln M}$  	 &  0.111 	 	 &  0.106 	 	 &  0.127 	 	 &  0.132 	 	 &  0.163 	 	 \\
$\sigma_{\ln T_{mw}}$  	 &  0.069 	 	 &  0.066 	 	 &  0.077 	 	 &  0.082 	 	 &  0.102 	 	 \\

\hline
\\
\multicolumn{6}{c}{\w\ self--similar scaling ($\alpha_{Y_X}=1.5$)} \\
$\log C_{T_{mw}}$ 	 &  14.413 $\pm$  0.007 	 & 14.410 $\pm$  0.005 	 & 14.414 $\pm$  0.005 	 & 14.417 $\pm$  0.004 	 & 14.419 $\pm$  0.005 	 \\
$\sigma_{\ln M}$	 &  0.134 	 	 &  0.127 	 	 &  0.147 	 	 &  0.142 	 	 &  0.168 	 	 \\
$\sigma_{\ln T_{mw}}$ 	 &  0.089 	 	 &  0.085 	 	 &  0.098 	 	 &  0.094 	 	 &  0.112 	 	 \\

\hline
\end{tabular}
\end{center}
\caption{The same as in Table \protect\ref{tab:YX}, but for the
  \mt\ relation. In this case we use $T_{\rm mw,0}=3\,{\rm
    keV}$ for the normalization factor in eq.(\protect\ref{eq:M_X}).}
\label{tab:Tmw}
\end{table*}

The deviations with respect to self--similarity that we detect in the
\mmg\ and \mt\ relations for radiative simulations are not
present in the \myx\ relation. In fact the best fit slope of
cluster data is $0.591\pm0.003$, close to the value of
$\alpha_{SS}=0.6$ predicted by the self--similar model. As apparent
from the right panel of Fig.~\ref{fig:SRred0}, the effect of including
cooling, star formation and SN feedback in the \w\ runs has only a
mo\-dest effect of the \myx\ scaling. The reason for this is that
the \yx\ mass proxy provides a measure of the total thermal content of
the ICM. Since this total thermal content is dominated by the
mechanism of gravitational accretion of baryons within the
DM--dominated potential wells of clusters, it is not surprising that
the scaling of \yx\ with total cluster mass very close to the
self--similar prediction and that its value, at a fixed cluster mass,
is weakly sensitive on the inclusion of non--radiative physics. This
is even more true when we exclude central cluster regions, as we did
in our analysis, which are more affected by radiative cooling.

Besides the stability of the scaling relations against variations of
the ICM physics, another useful quantity to judge the robustness of
mass proxies is the intrinsic scatter around the best fitting
relations. In fact, the intrinsic scatter and its distribution are
also an important quantities to look at when deciding which mass proxy
is the best one for cosmological applications of galaxy clusters
\citep[e.g.][]{Lima_Hu2005PhRvD..72d3006L,Shaw2010ApJ...716..281S}. Looking at
Fig.~\ref{fig:SRred0} we note that the mass--weighted temperature is
the proxy with the largest scatter, both for \w\ and \nrsv\ runs. To
display graphically the scatter of each relation we plot in 
Fig.~\ref{fig:scat_z0} 
the distribution of residuals, with respect to the
best--fit relation, in log--mass. In all panels the distribution of
the distribution of the residuals is close to a log--normal one.
The intrinsic mass scatter in the residuals of the $M$--\mg\ relation
is $\sigma_{\ln M}\simeq 0.039$ and $0.055$ for the \w\ and \nrsv\
simulations, respectively (see also Table \ref{tab:Mgas}). The largest
scatter is obtained for the mass--weighted temperature $T_{mw}$,
reaching values of $\sigma_{\ln M}\simeq 0.11$ and 0.12 for
non--radiative and radiative simulations, respectively, thus similar
to what found by \cite{Stanek2010} for their non--radiative
simulations (see also Table \ref{tab:Tmw}). As for the \yx\ proxy, the
intrinsic scatter in mass is $\sigma_{\ln M}\simeq 0.050$ and $0.064$
for \w\ and \nrsv\ simulations, respectively.

\begin{figure*}
\hbox{
\hspace{0.5truecm}
\psfig{figure=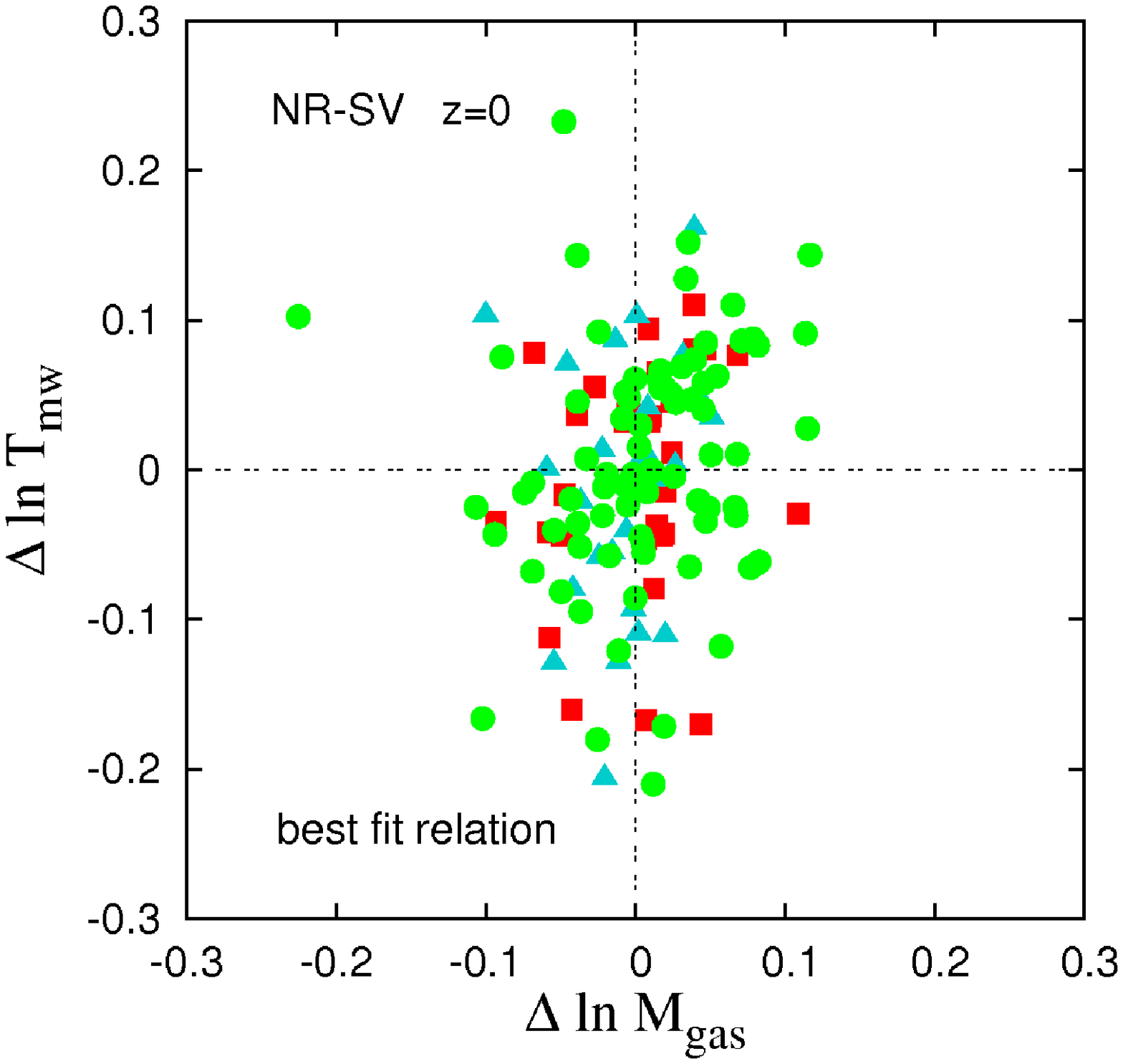,width=8.5cm,angle=-90}
\hspace{-0.truecm}
\psfig{figure=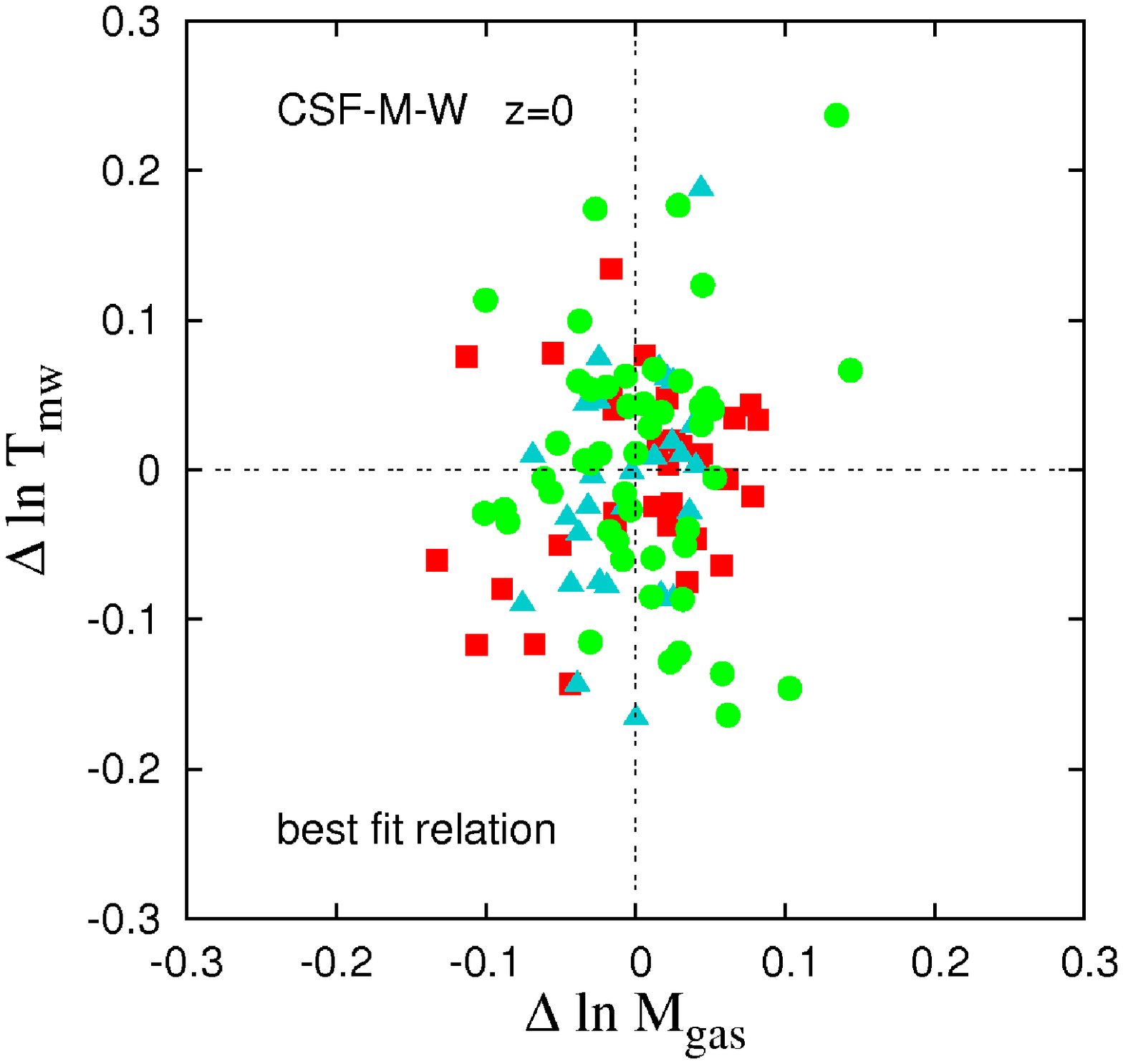,width=8.5cm,angle=-90}
}
\caption{The residuals in $\ln T_{mw}$ versus the residuals in $\ln
  M_{\rm gas}$ at fixed values of $M_{500}$ for clusters at $z=0$.
  Residuals are evaluated with respect to the best--fitting
  \protect\mt\ and \protect\mmg\ relations at $z=0$. 
  Left and right panels are respectively for non--radiative (\protect\nrsv) and
  radiative (\protect\w) simulations of clusters of Set 1. Different colours
  of the symbols correspond to different ranges in $M_{500}$. 
  Cyan triangles: $M_{500}>5\times 10^{14}M_\odot$;
  Red squares: $10^{14}<M_{500}<5\times 10^{14}M_\odot$;
  Green circles: $5\times 10^{13}<M_{500}<10^{14}M_\odot$.}
\label{fig:dev}
\end{figure*}

Therefore, while the \yx\ mass proxy is found to have a scaling with
mass which is closest to the self--similar behaviour, almost
independently of the inclusion of radiative physics and SN feedback,
the proxy having the lowest scatter against mass is \mg. This result
is in line with that obtained by \cite{Stanek2010} from the analysis
of clusters simulated with SPH, with both non--radiative physics and
including cooling plus the effect of a phenomenological
pre-heating. Similarly to our analysis, Stanek et al. did not attempt
to include observational effects in their analysis, while they
analysed scaling relations at a larger radius, $R_{200}$ instead of
$R_{500}$ like in our analysis. We also note that indications for a
lower scatter in the \mmg\ relation than in the \myx\ relation
have been also reported by \cite{Okabe2010arXiv1007.3816O} from the
analysis of observational data. Their analysis was based on
XMM--Newton data for the measurement of X--ray mass proxies and on
weak lensing data for the measurement of total cluster masses. These
results are however at variance with the original result by
\cite{Kravtsov2006ApJ...650..128K} who found instead \yx\ to be the
lowest--scatter mass proxy from their analysis of cluster simulations
based on the ART adaptive-mesh refinement code. In their analysis,
Kravtsov et al. showed that the residuals $\Delta T$ in the \mt\
relation and $\Delta $ \mg\ in the \mmg\ relation anticorrelate
for their radiative simulations when computed with respect to the
best--fitting self--similar relation, thus justifying the smaller
scatter found in the $M$--\yx\ scaling. Indeed the latter can be
computed as $\sigma^2_Y = \sigma^2_{T_{mw}} + \sigma^2_{M_{gas}} + 2
\; C_{T,M} \sigma_T \; \sigma_{M_{gas}}$, where $C_{T,M}$ is the
correlation coefficient and $\sigma$ the scatter values of temperature
and gas mass \citep{Stanek2010}.

\begin{figure*}
\hbox{
\hspace{0.5truecm}
\psfig{figure=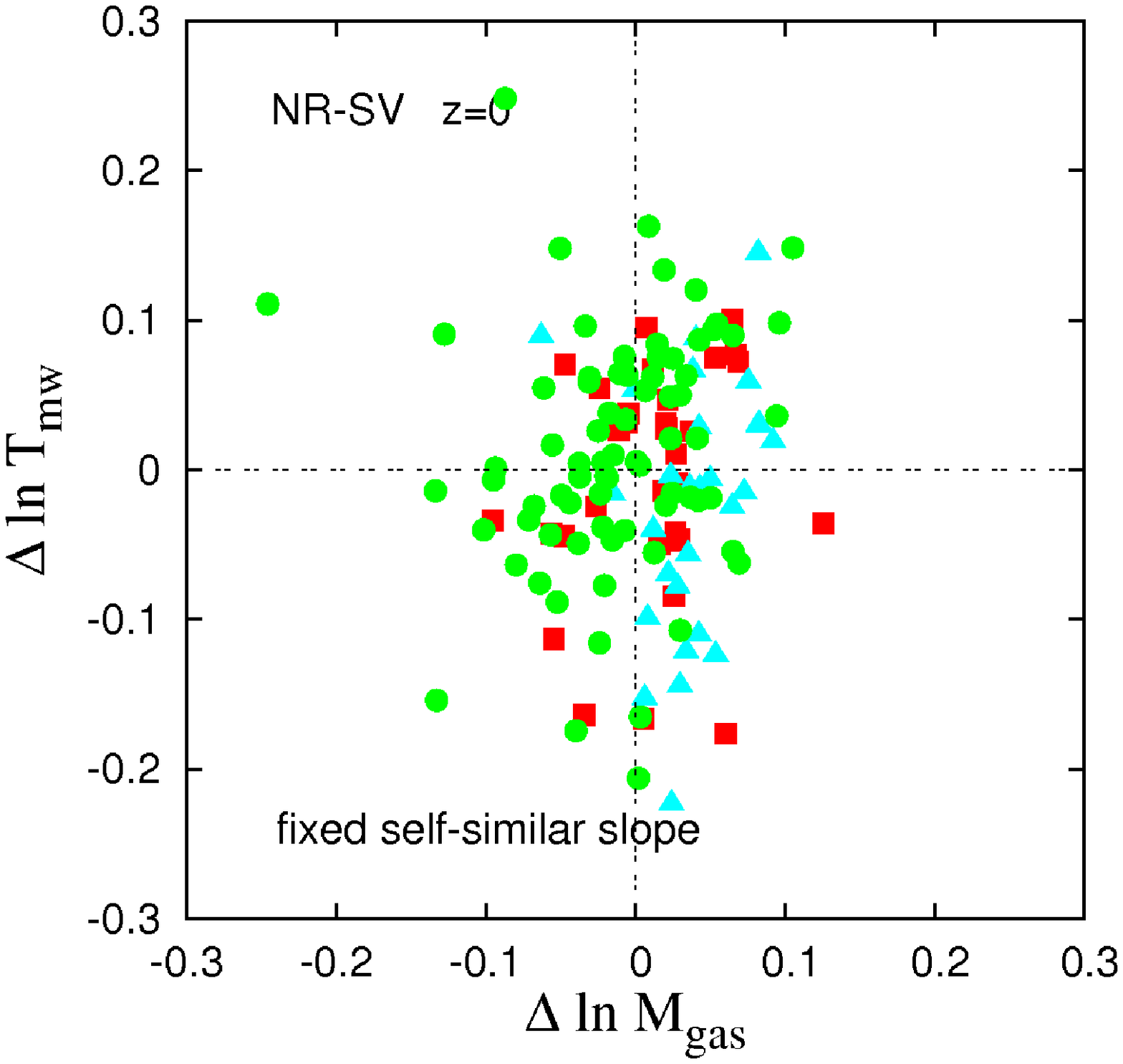,width=8.5cm,angle=-90}
\hspace{-0.truecm}
\psfig{figure=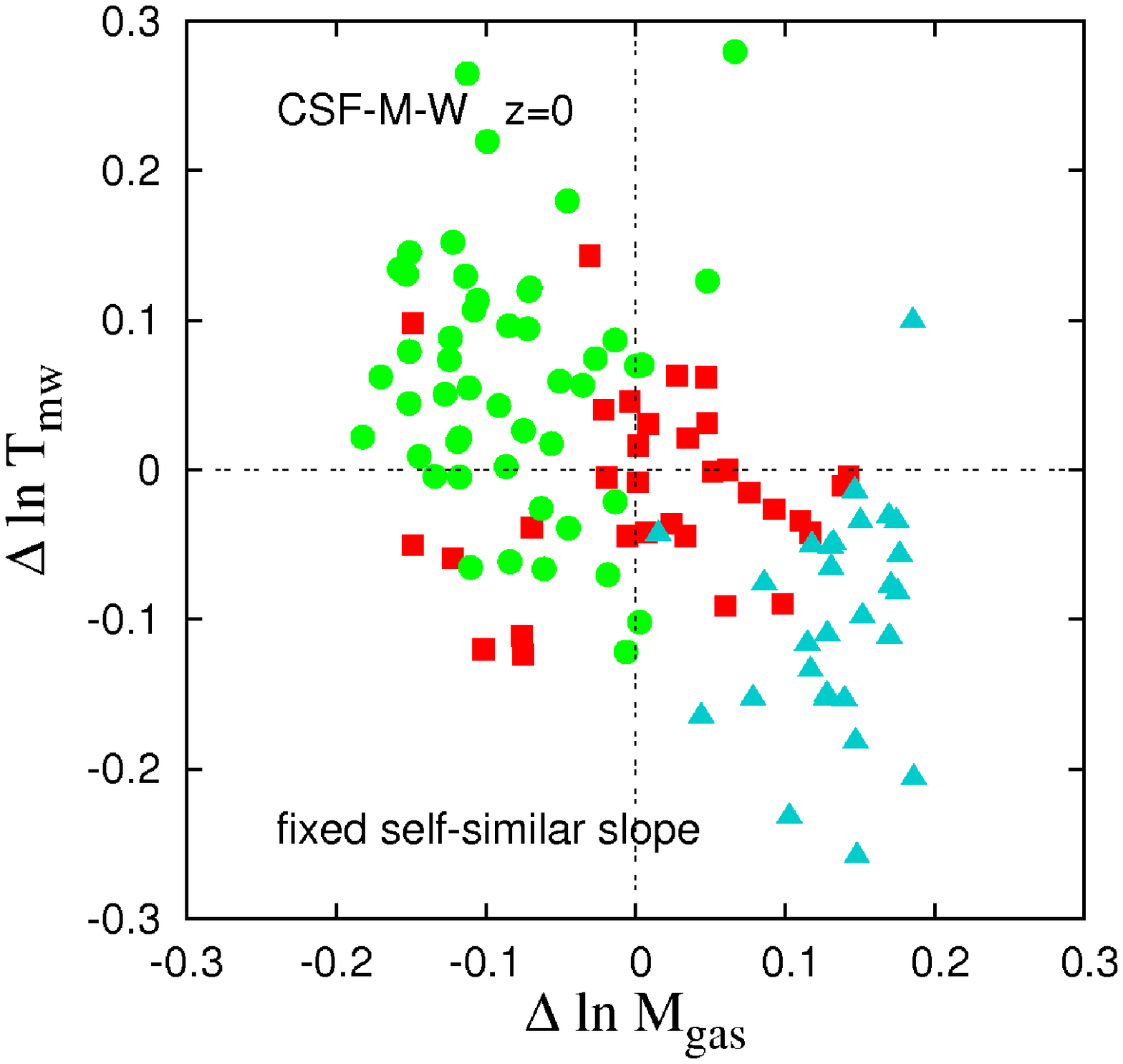,width=8.5cm,angle=-90}
}
\caption{The same as in Fig.~\protect\ref{fig:dev} but for the
  deviations with respect to the best--fitting relation obtained by
  fixing the slopes to the self--similar values. Colours of the
  symbols have the same meaning as in Fig.~\protect\ref{fig:dev}.}
\label{fig:dev_ss}
\end{figure*}

In order to look at the behaviour of such residuals in our simulations,
we plot in Fig.~\ref{fig:dev}
the residuals in $\ln \tmw$ versus the residuals in $\ln \mg$
determined at fixed total mass $M_{\rm 500}$. The residuals are computed
with respect to the best--fitting relations computed without imposing
the self--similar slope. 
Evaluating the Pearson's correlation coefficient $r_P$ for the residuals 
we found them to be uncorrelated. 
We find $r_P=0.24$ for \nrsv\ distribution (left panel) and
$r_P=0.11$ for the \w\ one (right panel).
The presence of a weak positive correlation explains why
the scatter on $Y_X$ is, if any, slightly larger than that on \mg. In
order to reproduce more closely the procedure followed by
\cite{Kravtsov2006ApJ...650..128K}, we then repeated the analysis by
computing the residuals with respect the best--fitting self--similar
relation. The results in this case are shown in Fig.~\ref{fig:dev_ss}. 
As for the non--radiative runs (left panel) we find
that $\Delta \ln \tmw$ and $\Delta \ln \mg$ are again consistent with
being uncorrelated for the \nrsv\ simulations, with $r_P =
0.14$. However, such scatters show a more evident sign of
anti--correlation for the \w\ simulations, $r_P = -0.53$. 

In order to make more clear the reason for the anti-correlation in the
residuals computed with respect to the self--similar best--fitting
relation for the radiative runs, we show with different symbols
in Figs.~\ref{fig:dev} and
\ref{fig:dev_ss} the points corresponding to clusters within different
mass ranges: 
the cyan triangles show the most massive objects with $M_{500} > 5\times
10^{14}M_\odot$, the green points refer to groups with $M_{500} <
10^{14}M_\odot$, while red squares represent the intermediate clusters.
The only panel showing anti--correlation, the right panel of
Fig.~\ref{fig:dev_ss}, also shows a clear trend with cluster mass.
Anti-correlation for \w\ clusters is actually a direct result of the
offset of scaling relations in temperature and gas mass with respect
to the expected self--similar ones. As already discussed, in the
radiative runs the two relations show an opposite behaviour with
respect to self-similarity: $M$--\mg\ and $M$--\tmw\ are
correspondinly shallower and steeper than the self--similar
scaling. This means that massive clusters in \w\ simulations do have
higher gas masses and lower temperatures. On the other side, in low
mass clusters the amount of gas is lower and temperatures are higher
than expected. Therefore, in this case the presence of an
anti-correlation is only a spurious effect of imposing the
self--similar slopes to the best--fitting scaling relations used to
compute such residuals. Clearly, in comparing our results with those
by \cite{Kravtsov2006ApJ...650..128K} one should remember that their
analysis attempted to include observational effects in the estimate of
the temperature and of the gas mass. The analysis presented here,
instead, is aimed at quantifying the intrisic performances of the
different mass proxies, i.e. before convolving with the effect of
measuring temperature from X--ray spectra and gas mass from an
projected map of X--ray surface brightness. Quite likely, including
such observational effects could affect both scaling relations and the
intrisinc scatter around them through the distribution of the
residuals. We will address in detail these issues in a forthcoming
paper (Rasia et al., in preparation). 

In summary, the results presented in this section have shown that:
{\it (i)} \mg\ has an intrinsic scatter against cluster mass
slightly smaller than \yx, while both such proxies have a
significantly lower scatter than \tmw; {\it (ii)} \yx\ is the proxy
whose scaling against mass has the weakest dependence on the inclusion
of radiative physics and whose slope is closest to the prediction of
the self--similar model. The latter result is in line with the
expectation that \yx\ is a measure of the total thermal energy content
of the ICM, which is mostly determined by gravitational processes once
central cluster regions are excluded from the analysis. Moreover, we 
found that residuals in \tmw\ and \mg\ with respect to the best--fit relations 
are uncorrelated independently of the adopted simulation, 
while anti--correlation seen in radiative 
simulation is as a result of the offset of \mt\ and \mmg\ relations 
with respect to the expected self--similar ones.
 
\subsection{Stability against changing ICM physics}
\label{SRphys}

In order to further investigate the sensitivity of the different mass
proxies on the physics included in the simulation, in this section we
use results from the analysis of the Set 2 simulated clusters. As
already discussed in Sect.~2, this set is cha\-rac\-te\-rized by a much
smaller number of simulated clusters than Set~1, but includes 7
different prescriptions to describe the baryonic physics. 

\begin{figure*}
\hbox{
\hspace{-0.2truecm}
\psfig{figure=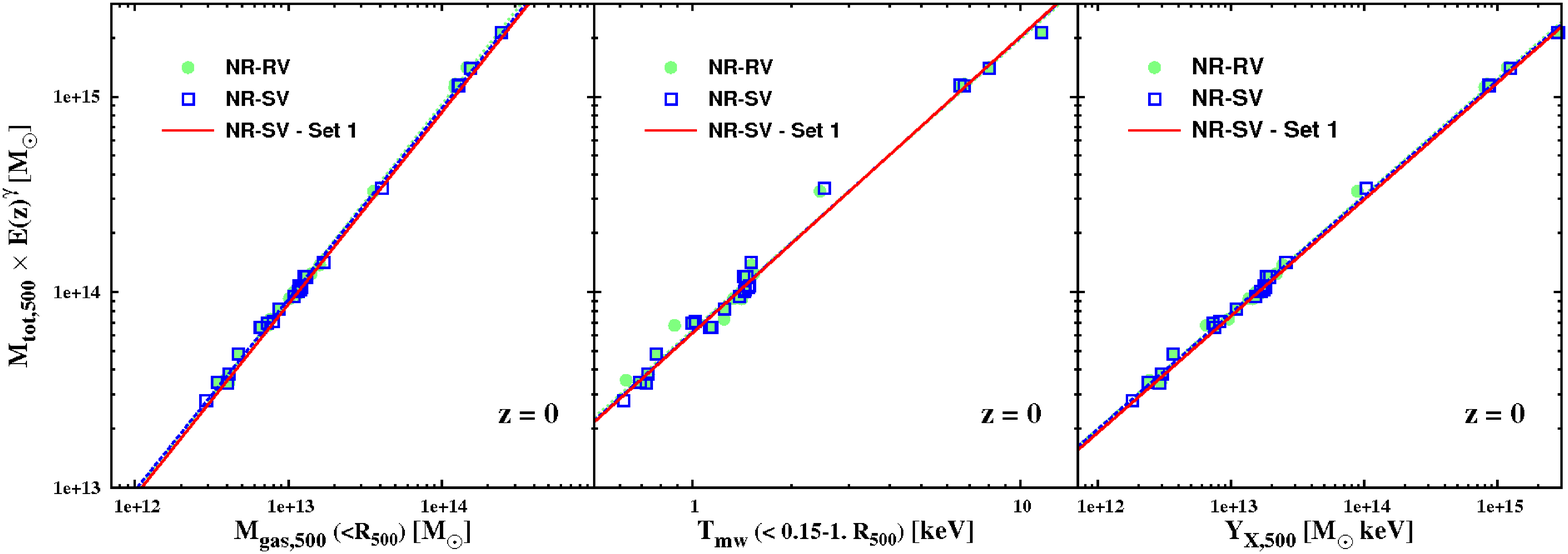,width=18.0cm,angle=0}
}
\caption{Scaling relations at $z=0$ between the total mass and three
  mass proxies (\protect\yx\ and \protect\mg\ computed inside $R_{\rm 500}$, 
  \protect\tmw\
  evaluated extracting the central $0.15 R_{\rm 500}$). 
  The evolution of the relation is scaled accordingly, with
  $\gamma = 2/5, 0$ and $1$, respectively.
  Best fits of
  non--radiative Set~2 simulations are represented by long dashed
  green and short dashed blue lines (\protect\nrrv\ and \protect\nrsv\ 
  respectively), compared with the best fit of the Set~1 
  \protect\nrsv\ clusters with red continuous line. 
  For this comparison we scaled the results of \protect\mg\ and 
  \protect\tmw\ obtained for clusters of Set~1 simulations 
  to the baryon fraction of Set~2.}
\label{fig:compNR}
\end{figure*}

As first test we verify that simulated clusters from Set~1 and Set~2
do produce consistent results when simulated with the same physics
(despite the fact they are based on somewhat different cosmological
models). We show in Fig.~\ref{fig:compNR} results from the
non--radiative runs (\nrsv\ and \nrrv) of Set~2 and compare them with
the best--fitting relation for the \nrsv\ version of Set~1 (continuous
line). In performing this comparison, we should account for the
different values of the baryon fraction assumed in the two simulation
sets (see Sect. 2.1). To account for this, we scaled \mg\ and
\yx\ for the Set 1 simulation \nrsv\ by multiplying them by the ratio
of the two baryon fractions: $f_{b,Set 1}/f_{b,Set 2} \simeq =1.2$.
From this plot we note that results for the \nrsv\ versions of Set 1
and Set 2 agree very well with each other. Furthermore, we also note
that the best fitting relations for the \nrsv\ and \nrrv\ simulations
of clusters of Set 2 are also extremely similar. As shown by
\cite{Dolag2005MNRAS.364..753D}, using the time--dependent reduced
viscosity scheme has the effect of increasing turbulent motions in the
ICM \citep[see also][]{Valda2010arXiv1010.3378V}. Therefore, the
stability of results against variations of the adopted viscosity
scheme (still within SPH hydrodynamics) implies that non--thermal
pressure support associated to turbulent motions have only a very
modest impact on the overall gas mass and thermal content in our
simulated clusters.

\begin{figure*}
\hbox{
\psfig{figure=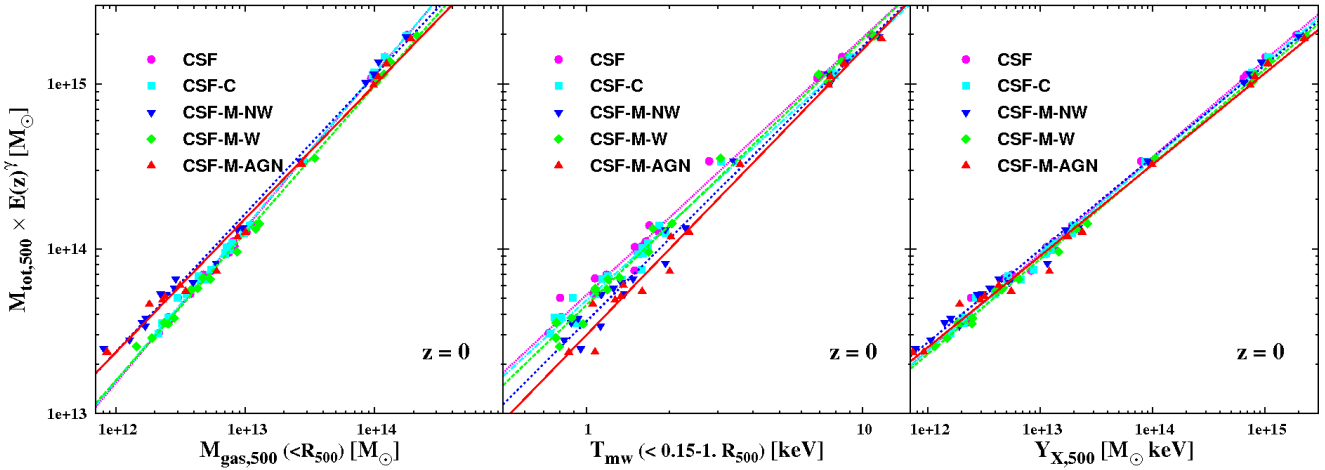,width=18.0cm,angle=0}
}
\caption{Effect of different physics models on scaling relations
  between total mass and \protect\yx, \protect\mg\ and \protect\tmw\ 
  at $z=0$ for the
  radiative simulations of Set 1. 
  The evolution of the relation is scaled accordingly, with
  $\gamma = 2/5, 0$ and $1$, respectively. For each simulated physics we plot
  cluster data with purple dots (\protect\csf), cyan squares 
  (\protect\csfc), blue triangles (\protect\nw), green diamonds 
  (\protect\w) and red triangles (\protect\agn). For
  clarity, data points are overplotted with the best-fitting relation
  for each of the five physics, using the same colour coding of the
  points.}
\label{fig:phys}
\end{figure*}

As for radiative simulations, we verified the effect of chan\-ging
feedback efficiency and of including thermal conduction. All such
processes are expected to change the gas distribution and the
temperature structure of the ICM and, therefore, to potentially impact
on the scaling relations of mass proxies against total cluster mass. 
We plot in Fig.~\ref{fig:phys} 
such scaling relations obtained the different radiative runs, \csf, \csfc, \nw, \w\ and 
\agn\, along with their best--fitting relations. 

The relations involving gas mass and temperature (left and central
panel, respectively) show significant dependencies on the physical
processes included in the simulations, especially for relatively low
mass systems.  In general, while the effect of including thermal
conduction is quite small, the separation between simulations with and
without AGN feedback is more evident. Indeed, AGN feedback has the
effect of removing a significant amount of gas from the innermost
regions \citep[e.g.,][]{Fabjan2010MNRAS.401.1670F}, thus inducing a
decrease of \mg, which is more pronounced for galaxy groups than
for rich clusters.  This effect can be observed in the central panel
of Fig.~\ref{fig:phys}, where the slope of the best fit relation for
AGN simulations is of $\alpha=0.81$, thus significantly flatter than
the $\alpha=1$ self--similar scaling. The effect of AGN is not only to
remove a big amount of gas from the center of galaxy groups but also
to heat up the ICM in the central regions of such systems. This can be
seen in the \mt\ relation shown in the central panel,
where the resulting slope of the relation is $\alpha=1.73$, again
quite different from the $\alpha=1.5$ self--similar value.
Simulations including the effect of galactic winds, namely \csf,
\csfc\ and \w, do have a slope closer to the self--similar one for
both the \mmg\ ($\alpha \simeq 0.9$) and the
\mt\ ($\alpha \simeq 1.5$--1.6) relation. The clusters
simulated without feedback have an intermediate behaviour for both
relations. Similarly to what already seen for simulations of Set 1,
\mt\ is the scaling relation characterized by the largest
scatter. 

\begin{figure*}
\hbox{
\psfig{figure=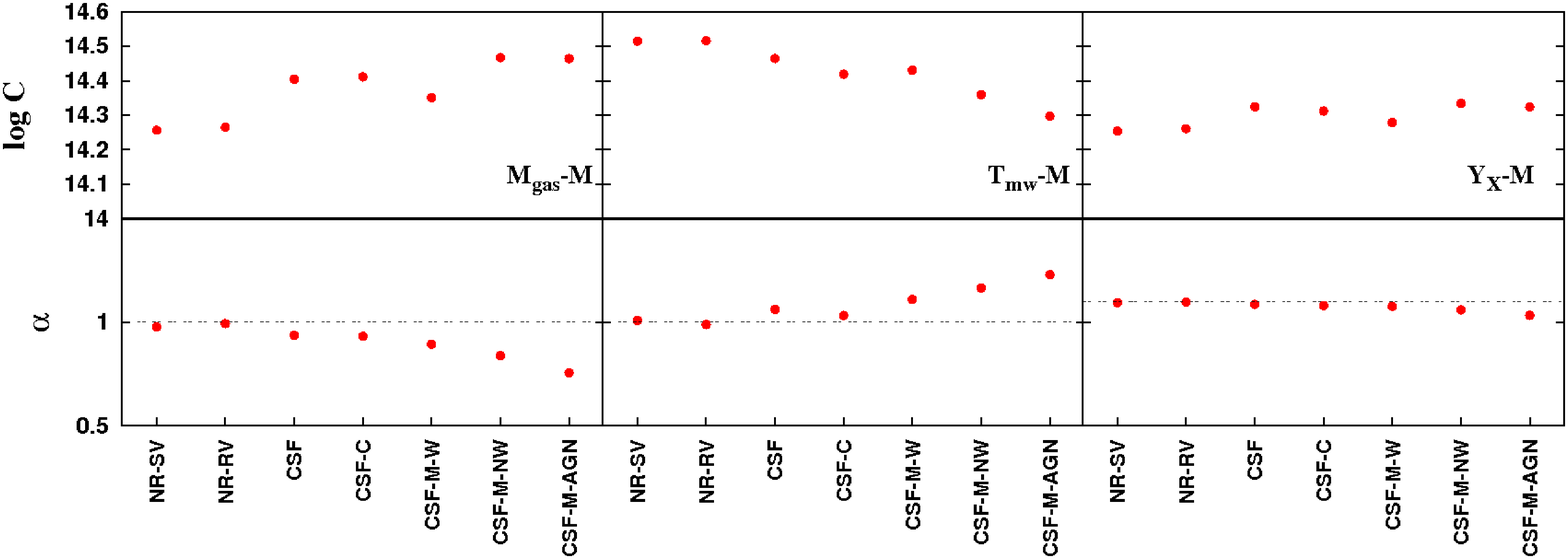,width=18.0cm,angle=0}
}
\caption{Dependence of the best--fitting slope $\alpha$ (lower panels)
  and normalization $C$ (upper panels) of the scaling relations of the
  three mass proxies against total cluster mass. Results for for
  \mmg, \mt\ and \myx\ are
  shown in the left, central, and right panel, respectively. In the
  lower panel, the horizontal dashed lines mark the values of the
  slope of the scaling relations predicted by the self--similar model.}
\label{fig:models}
\end{figure*}

On the other hand, the \myx\ relation (right panel in Fig.~\ref{fig:phys}) is
confirmed to be the one with the weakest sensitivity on the baryon
physics included in the simulation. To further emphasize the different
sensitivity that different mass proxies have on the ICM physics, we
plot in Fig.~\ref{fig:models} the logarithm of the normalization,
$\log C$ and the slope $\alpha$ of all the best fit scaling relations
for the seven different physical models simulated for the clusters of
Set 2. \tmw\ and \mg\ show the largest discrepancies in $\log C$
varying from $14.2$ to $14.5$ in both relation. The \yx\ proxy has
instead a more stable behaviour with values around
$14.25--14.35$. Radiative runs show also a consistent off-set from the
self--similar slopes in the case of \mg\ and \tmw\ while the slope of
\myx\ relation is in good agreement with the self-similar
value of $\alpha=0.6$ for all the models considered. A maximum
deviation of $\simeq 11\%$ is only present in the case of \agn\
feedback.

In summary, in this section we have shown that the \yx\ mass proxy is
the most robust one against changing the physical processes which
determine the thermodynamical properties of the ICM and its scaling
with total cluster mass has a slope which is always very close to that
predicted by the self--similar model.

\subsection{Evolution of scaling relations}
\label{Sec:evol}

For mass proxies to be used in cosmological applications of galaxy
clusters, one has to precisely calibrate not only normalization, slope
and scatter of their scaling relations with total mass for
low--redshift systems, but also their evolution with
redshift. Redshift evolution is especially important when considering
distant clusters, located at the highest redshifts, $z\magcir 1$,
where they have been secured to date in statistically complete X--ray
surveys. Indeed, as we approach the epoch of cluster assembly one
expects clusters to be characterized by major mergers, which may
significantly impact on the shaping of such scaling relations. This
issue is of crucial importance if we want to constrain cosmological
parameters by using upcoming and future large cluster surveys, based
on the next generation of wide field X--ray telescopes
\citep[e.g.,][]{Borgani2010arXiv1010.6213B,Cappelluti2010arXiv1004.5219C}. 

In this Section we will address the evolution of scaling relations
and of their intrinsic scatter. For an accurate study of the
evolution of scaling relations we resort to the large statistical
database of clusters for Set 1 simulations. All simulated clusters are
analysed at five redshifts: $z=0$, $0.25$, $0.5$, $0.8$ and $1$. After
measuring the normalization $C$ and slope $\alpha$ of the scaling
relations at each redshift, their redshift evolution is described
through the relations:
\ba
\log C (z) & = & \log C_0(z) + \beta_1 \left( 1 + z \right) ;\nonumber \\
\alpha(z) & = & \alpha_0 + \beta_2 \left( 1 + z \right) \, .
\label{eq:fit2}
\ea 
Similarly to what done by \cite{Short2010arXiv1002.4539S}, who studied
the evolution of scaling relations from Millenium simulations, we take
the redshift dependence of $C_0(z)$ to be that predicted by the
self--similar model. Therefore, we expect $\beta_1=0$ for a
self--similar evolution of the normalization.

\begin{table*}
\begin{center}
\begin{tabular}{lcccc}
\hline
 & $\log C_0$	& $\beta_1$ & $\alpha_0$ & $\beta_2$ \\ 
\hline
\hline
\\
\multicolumn{5}{c}{$M_{\rm gas}$--$M_{\rm 500}$} \\
\w\		& $14.307 \pm 0.003$ & $-0.079 \pm 0.015$ 
		& $0.927 \pm 0.003$ & $0.005 \pm 0.017$ \\
\nrsv\		& $14.152 \pm 0.001$ & $-0.086 \pm 0.007$ 
		& $0.979 \pm 0.002$ & $0.057 \pm 0.011$ \\
\hline
\\
\multicolumn{5}{c}{$T_{\rm mw}$--$M_{\rm 500}$} \\
\w\           & $14.430 \pm 0.005$ & $0.059 \pm 0.025$ 
		& $1.622 \pm 0.014$ & $-0.047 \pm 0.070$ \\
\nrsv\		& $14.518 \pm 0.011$ & $0.020 \pm 0.054$ 
		& $1.524 \pm 0.012$ & $-0.075 \pm 0.061$ \\
\hline
\\
\multicolumn{5}{c}{\yx--$M_{\rm 500}$} \\
\w\		& $14.248 \pm 0.002$ & $0.007 \pm 0.008$ 
		& $0.591 \pm 0.002$ & $0.011 \pm 0.012$ \\
\nrsv\		& $14.191 \pm 0.004$ & $-0.030 \pm 0.019$ 
		& $0.597 \pm 0.002$ & $0.023 \pm 0.009$ \\
\hline
\end{tabular}
\end{center}
\caption{Best fitting values for the parameters determining the
  evolution of 
  normalization $C$ and slope $\alpha$ of the scaling relations of
  \mg, \tmw\ and \yx\ against $M_{\rm 500}$, as described by  
  eq.(\ref{eq:fit2}), over the redshift range $0\le z\le 1$. For each
  relation the values obtained for both \protect\w\ and \protect\nrsv\ 
  simulations are reported. Reported uncertainties correspond to 1$\sigma$
  uncertainties in the linear regression. 
}
\label{tab:evol}
\end{table*}

\begin{figure*}
\hbox{
\psfig{figure=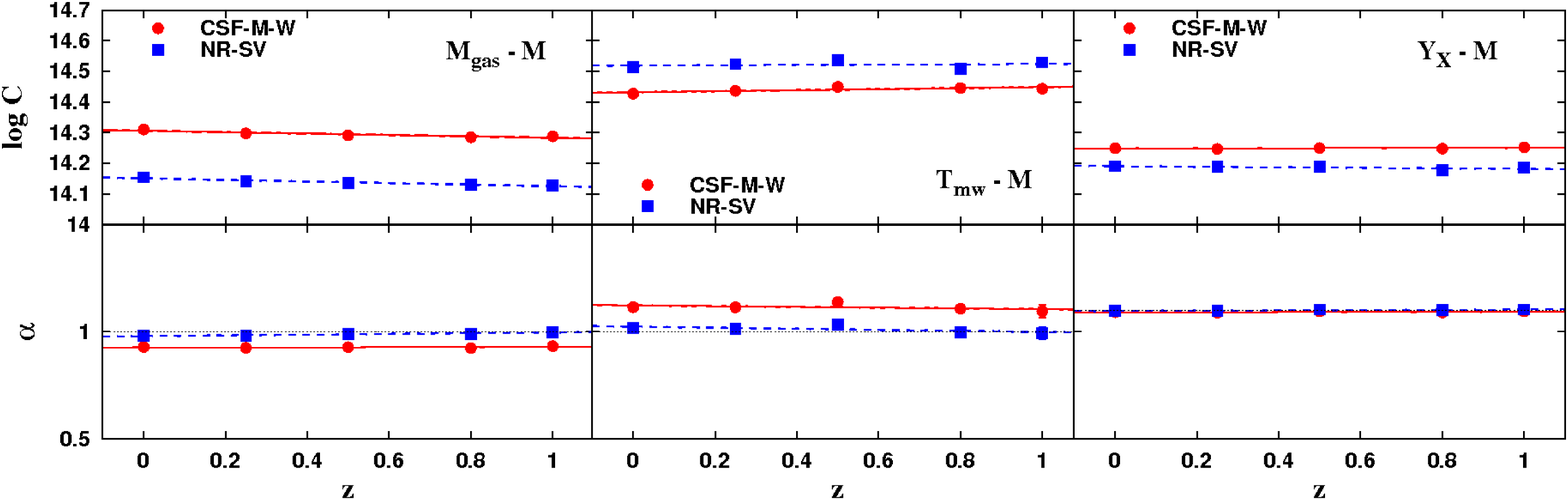,width=18.0cm,angle=0}}
\caption{Redshift--dependence of the best fitting parameters which
  define the scaling relations between mass proxies and
  $M_{\rm tot,500}$. Left panel: scaling with $Y_{\rm X,500}$; 
  central panel: scaling with $M_{\rm gas,500}$; right panel: scaling with
  mass-weighted temperature, $T_{\rm mw}$ (see
  eq.\protect\ref{eq:fit2}). Results for \protect\w\ and 
  \protect\nrsv\ simulations 
  are plotted with red circles and blue squares, respectively.
  For each point we also plot the errorbars
  corresponding to 1$\sigma$ uncertainty from the log--log linear
  regression, whose size is smaller than the size of the points. The
  horizontal dotted lines in lower panels mark the values of the slope $\alpha$
  predicted by the self--similar scaling. As for the amplitude $C$,
  its redshift dependence mark the degree of deviation from the
  evolution predicted by the self--similar model.}
\label{fig:fitpar_1}
\end{figure*}

We list in Table \ref{tab:evol} the values of the parameters which
describe through eqs.\ref{eq:fit2} the evolution of the scaling
relations between the three mass proxies \mg, \tmw\ and \yx\ with the
cluster total mass. Moreover, we plot in Fig.~\ref{fig:fitpar_1} the
redshift dependence of normalization (after accounting for the
self--similar evolution; upper panels) and slope (lower panels) of
these scaling relations. The \mg\ proxy has a very mild negative
evolution with redshift ($\beta_1 < 0$) for both radiative and
non--radiative simulations. The slope is confirmed to be very close to
the self--similar value for non--radiative simulations at all
redshift, with a negligible evolution of its value, while being
shallower for the radiative simulations. As for the mass--weighted
temperature, a slightly positive evolution is detected for the \csf\
simulations, while the evolution is very close to self--similar for
\nr\ simulations. 

Quite interestingly, the \myx\ scaling relation is the one
having both the weakest sensitivity to the ICM physics at all
redshifts and the smallest deviations from the self--similar
predictions on the evolution of the normalization and the value of the
slope. These results confirm once again the robustness of the \yx\
mass proxy against variations of the physics of the ICM over the
whole considered redshift range. They further confirm that, since this
mass proxy provides a measure of the ICM total thermal energy content,
it can be reliably described on the ground of the predictions of the
self--similar model. Indeed, this model is based on the assumption
that gravitational mechanisms determines the thermal content of the
ICM, an assumption which is expected not to be violated by radiative
and feedback processes, once we exclude core regions of galaxy
clusters.

\begin{figure*}
\hbox{
\hspace{-0.2truecm}
\psfig{figure=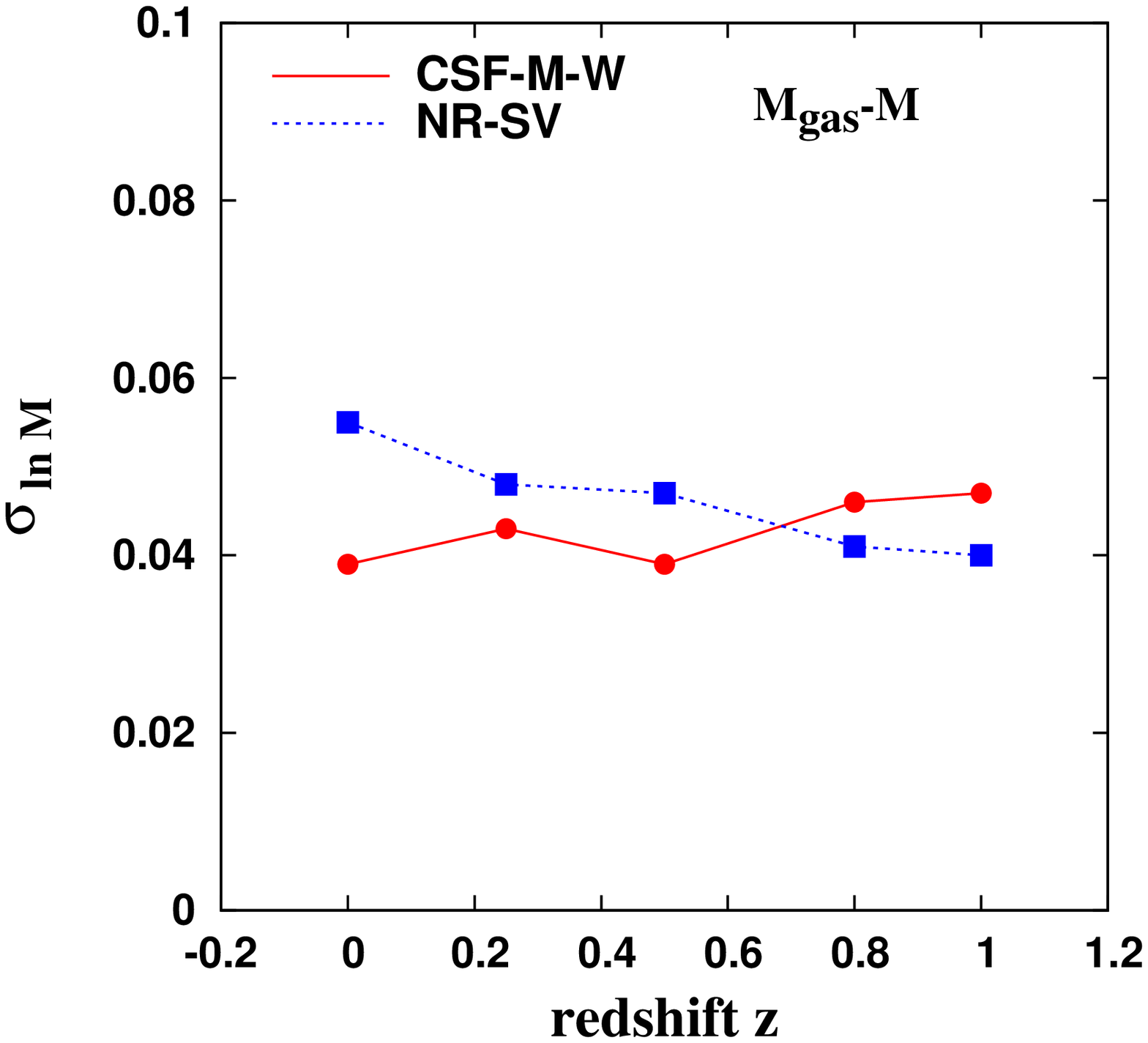,width=6.0cm,angle=-90}
\hspace{-0.2truecm}
\psfig{figure=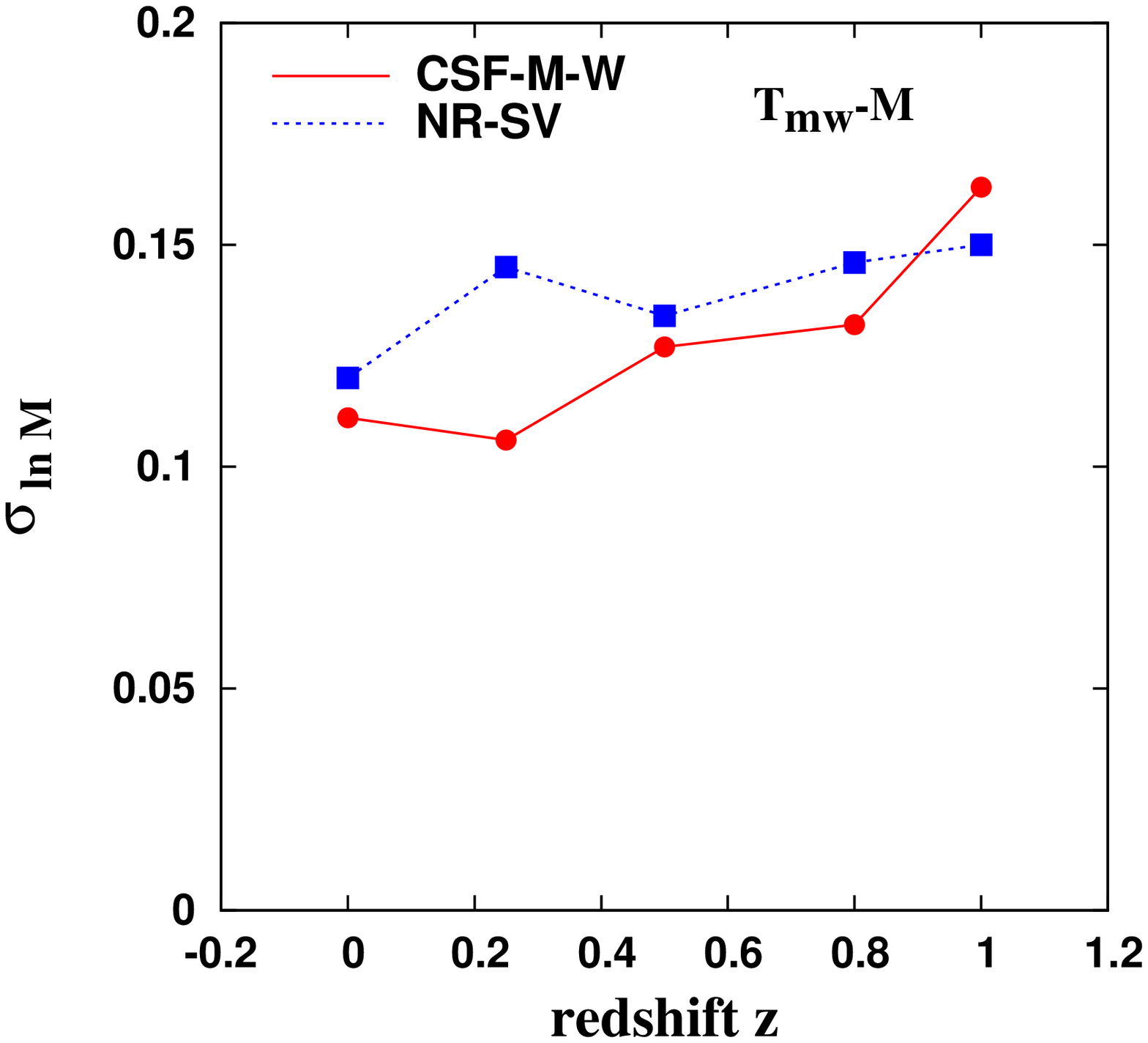,width=6.0cm,angle=-90}
\hspace{-0.2truecm}
\psfig{figure=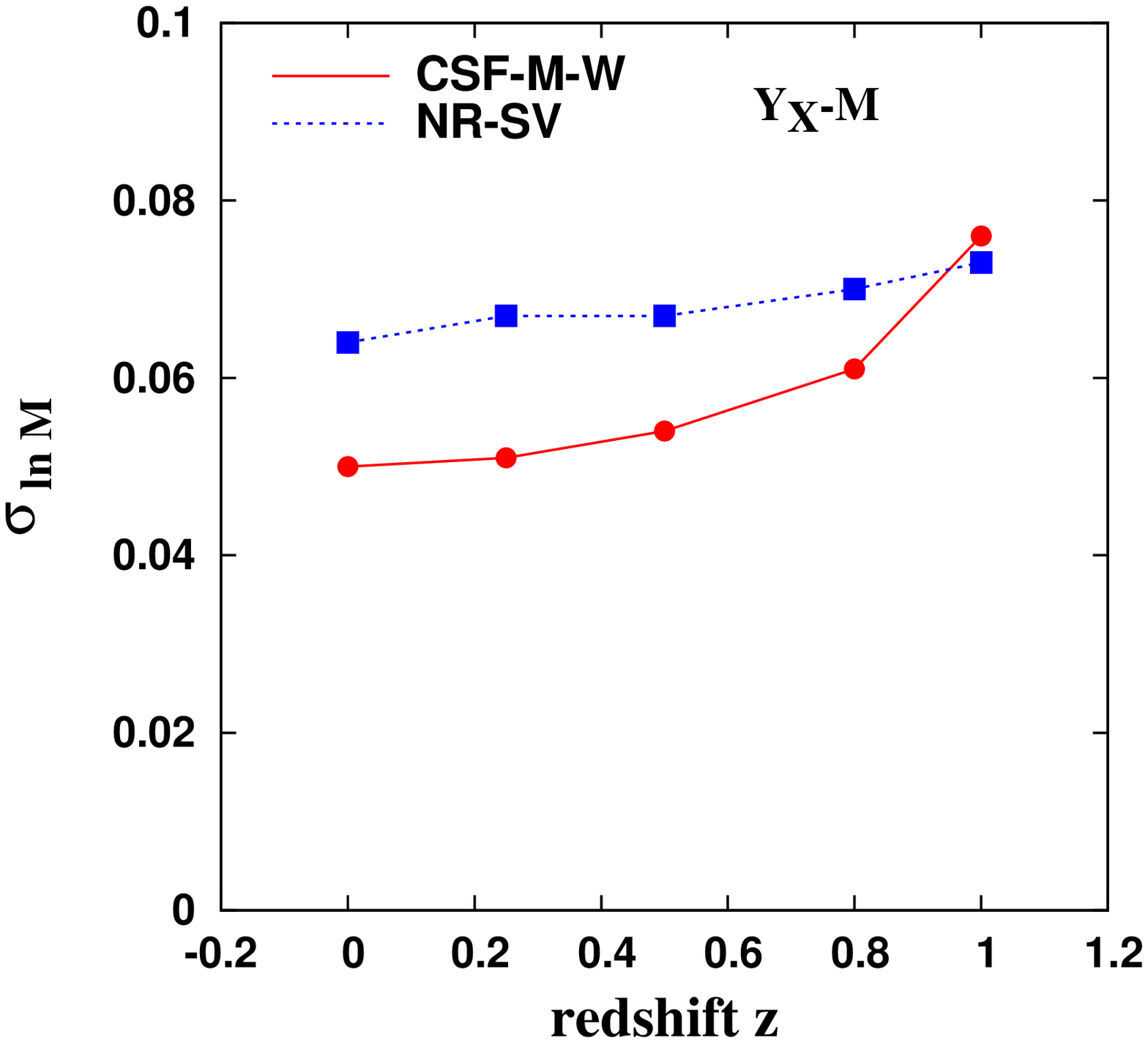,width=6.0cm,angle=-90}
}
\caption{Redshift evolution of the intrinsic scatter $\sigma_{\ln M}$
  in the scaling relations for simulated clusters of Set 1: results are 
  plotted with red circles for \protect\w\ simulations and blue squares for 
  \protect\nrsv\ simulations. Left,
  middle and right panels show results for the $M_{\rm tot,500}-M_{\rm
  gas}$, $M_{\rm tot,500}-Y_{X}$, and $M_{\rm tot,500}-T_{mw}$, respectively. 
  Note that the y--axis range in the middle panel is twice as large as
  for the other two panels.}
\label{fig:fitpar_2}
\end{figure*}

A further important aspect that characterizes the behaviour of the three
analysed mass proxies at different redshifts is the evolution of their
intrinsic scatter. We plot the evolution of these intrinsic scatters
in Fig.~\ref{fig:fitpar_2} for both \csf\ (red circles) and \nr\
(blue square) simulated clusters. The intrinsic scatter for the
\mmg\ relation, shown in the left panel of Fig.~\ref{fig:fitpar_2}, 
is remarkably low for both simulations, with a
slight positive (negative) evolution for the radiative (non--radiative)
simulations. In general, the intrinsic scatter in mass,
$\sigma_{\ln M}$, for \mg\ varies in the narrow range $4-5$
per cent at all considered redshifts. \csf\ clusters instead show a
mild decrease with a constant $4\%$ scatter below $z < 0.5$. Such
small values are expected, since gas mass is not sensitive to cluster
mergers, which are expected to play an increasingly important role at
higher redshift. The middle panel of Fig.~\ref{fig:fitpar_2} shows
instead the evolution of intrinsic scatter for the \mt\
relation. In this case we note that the intrinsic scatter is larger and
shows clear sign of a positive evolution for both radiative and
non--radiative simulations, with an increase from $\sigma_{\ln
  M}\magcir 10$ per cent at $z=0$ to $\magcir 15$ per cent at $z=1$.
The increase of the scatter in the \mt\ relation agrees
with the expectation that temperature is more sensitive than gas mass
to the presence of substructures in the ICM, which are expected to be
more prominent at high redshift, when cluster mergers are more
frequent. As for the \myx\ scaling relation, it has a
positive evolution which is driven by the positive evolution of the
scatter in the \mt\ relation (right panel of
Fig.~\ref{fig:fitpar_2}). In general the intrinsic scatter increases
from $\sigma_{\ln M}\simeq 5-6$ per cent at low redshift to $\mincir
8$ per cent at $z=1$. 

The analysis presented in this Section confirms that gas mass is the
mass proxy behaving best as for the intrinsic scatter in its scaling
relation with total mass, both for its value and for the stability
with redshift. 

\vspace{0.3truecm} An important {\it caveat} not to over interpret the
results presented here concern the fact that the small values of
intrinsic scatter for all mass proxies have been obtained by
neglecting observational effects in the measurement of mass
proxies. Such observational effects are in general expected to
introduce intrinsic scatter. A typical example is provided by the
presence of substructures which are not resolved in realistic
observational conditions. The contribution of the X--ray emissivity
from the high--density, low--temperature substructures impacts both on
the estimate of the gas mass from the X--ray surface brightness
profile and on the spectroscopic measurement of temperature
\citep[][]{Mazzotta2004MNRAS.354...10M,Vikhlinin2006ApJ...640..710V}. In
this way, observational determinations of gas mass and temperature
will have a larger scatter in their scaling relation with cluster mass
than their mass--weighted counterpart do.

\section{Conclusions}
\label{sec:conc}
Galaxy clusters are powerful tools for cosmological studies, since the
evolution of their mass function constrain the nor\-ma\-li\-za\-tion of the
power spectrum, the density parameter of dark matter and dark energy
as well as the dark energy equation of state, through the linear
growth rate of density perturbations. When observed in the X--ray
band, their high emissivity allows clusters to be detected so far out
to high redshifts, $z \sim 1.5$. However, to fully exploit the
potential of galaxy clusters as tracers of cosmic evolution, it is
necessary to understand in detail the relation between mass proxies,
based on easy-to-measure X--ray observables, and total cluster mass.

In this paper we focused on the study of the \mg\, \tmw\ and \yx\ mass
proxies and the effect that different physical mechanisms have on the
mass--observable relations. The aim of this analysis was to answer
through simulations to three questions: (1) Which is the mass proxy
that is least sensitive to the uncertain knowledge of the physical
processes determining the thermodynamical structure of the ICM? (2) To
what extent such mass proxies follow the prediction of the
self--similar model for the shape of scaling relations and their
redshift evolution? (3) How large is the intrinsic scatter in these
scaling relations and how does it evolve with redshift? 

To answer these questions we used galaxy clusters simulated with SPH
using the \gadget. We combined two different sets of simulations: Set~1, 
composed by more than hundred clusters with mass above $5 \times
10^{13} h^{-1} M_{\odot}$ was used for its high statistics to
calibrate scaling relations and study their scatter and redshift
evolution; Set~2, containing many fewer clusters, simulated with seven
different different physics schemes was instead used to study the
effect of: {\it (i)} thermal conduction, {\it (ii)} artificial
viscosity, {\it (iii)} cooling and star formation, {\it (iv)} galactic
winds and {\it (v)} AGN feedback.

The main results of our analysis can be summarised as follows.
\begin{description}
\item[(1)] In non--radiative simulations the relations between cluster
  total mass and the three considered mass proxies follow closely the
  self--similar prediction.
\item[(2)] In radiative simulations the \mmg\ and
  \mt\ sca\-ling relations show an opposite deviation from
  self--si\-mi\-la\-ri\-ty. The net result is a sort of compensating effect in
  \myx\ relation, with a nearly self--similar slope.
\item[(3)] The \myx\ relation is the most stable against
  changing of the physical processes included in the simulations, with
  its slope and evolution being always very close to the predictions
  of the self--similar model, which is based on the assumption that
  gravity only drives the ICM thermodynamics.  Indeed, the \yx\ proxy
  is by definition a measure of the total thermal content of the ICM,
  which is dominated by the gravitational process of gas accretion.
\item[(4)] \mmg\ is found to be the scaling relation with the lowest
  scatter in mass, $4-6$ per cent. Moreover, the amount of scatter in this case 
is almost constant with redshift, independently of the simulated physics. 
\item[(5)] The scatter in the \myx\ relation is slightly
  larger that for the \mmg\ relation; its intrinsic scatter
  grows with redshift as a consequence of the increase of merging
  events and presence of substructure in the ICM.
\end{description}

\begin{figure}
\psfig{figure=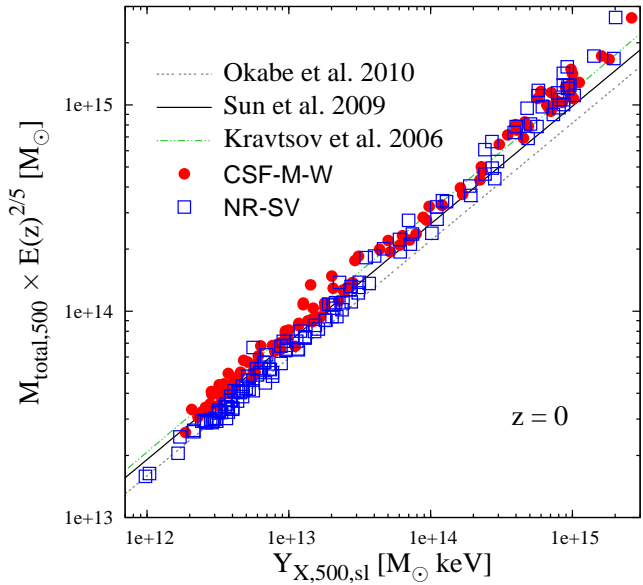,width=9.0cm,angle=-90}
\caption{Comparison between the $M_{\rm tot,500}$--$Y_{\rm X,500, sl}$ 
  relations for radiative simulations (\protect\w with red circles) 
  and non--radiative simulations (\protect\nrsv with blue squares) 
  of clusters of Set~1 at $z=0$, and results presented in the
  literature from simulations and observational data. For this
  comparison, we resorted to the spectroscopic--like definition of
  temperature
  \protect\citep{Mazzotta2004MNRAS.354...10M,Vikhlinin2006ApJ...640..710V}
  for the computation of $Y_{\rm X,sl}$ in our simulations. The dot--dashed
  (green) line corresponds to the best--fit relation from the
  radiative simulation by
  \protect\cite{Kravtsov2006ApJ...650..128K}. The continuous
  (black) line is the best fit relation found by
  \protect\cite{Sun2009ApJ...693.1142S} from the analysis of Chandra
  data of clusters and groups. The long--dashed (grey) line is the
  best--fit found by \protect\cite{Okabe2010arXiv1007.3816O} for the
  relation between $Y_X$ computed from XMM--Newton data and weak
  lensing masses from Subaru observations.}
\label{fig:MY_comp}
\end{figure}

As already discussed in previous sections, the analysis discussed in
this paper differs in spirit from that presented by
\cite{Kravtsov2006ApJ...650..128K}, who instead presented results
based on the inclusion of projection effects and spectroscopic
estimate of the ICM temperature. In order to have a first assessment
on the effect of carrying out a more observationally--oriented
analysis, we also computed the \yx\ mass proxy by using the
spectroscopic--like definition of temperature, $T_{\rm sl}$. To this
purpose, we followed the procedure described by
\cite{Vikhlinin2006ApJ...640..710V}, which generalizes the analytic
formula originally introduced by \cite{Mazzotta2004MNRAS.354...10M},
to include relatively cold clusters with temperature below 3 keV.
\footnote{We use the algorithm proposed by
  \cite{Vikhlinin2006ApJ...640..710V} that resorts to precomputed
  tables of some parameters for the observed spectra as a function of
  the temperature. Tables were generated with the code that is
  publicly available at {\tt
    http://hea-www.harvard.edu/~alexey/mixT}. The tables were created
  by fixing $N_H = 5 \, \times \, 10^{20}$ for the galactic hydrogen
  column density, solar abundances to the values by
  \cite{Grevesse1998SSRv...85..161G}, the Chandra ACIS-S CCD response
  function and the photon energy range to $0.7-10$ keV.}  We plot in
Fig.~\ref{fig:MY_comp} the scaling relation between total mass and
$Y_{\rm X,500,sl}=M_{\rm gas} \times T_{\rm sl}$ for the radiative 
and non--radiative versions of the clusters of Set~1. 
As expected the scatter in this
case is larger than when adopting the mass-weighted definition of
temperature: it increases
from $\sigma_{\ln M}=0.06$ to $0.10$ and from $\sigma_{\ln M}=0.05$ to $0.08$
for the \nrsv\ and \w\ simulations, respectively. Moreover also a
slight deviation from self--similarity is observed for non--radiative 
runs, with $\alpha=0.64$, while the slope for radiative runs, $\alpha=0.60$, 
agrees with the self--similar value. In the
same figure we also compare our results with those presented in
literature, both from simulations
\citep[][]{Kravtsov2006ApJ...650..128K} and from observational data
\citep[][]{Sun2009ApJ...693.1142S,Okabe2010arXiv1007.3816O}.  Our
results are in a good agreement although some discrepancy is seen at
the high mass end of clusters and for \nrsv\ galaxy groups.

Clearly, a more detailed analysis of how observational effects will affect
the scaling relations measured from our simulated cluster sets
requires using a dedicated software, like X--ray Map Simulator
\citep[X--MAS; e.g.][]{2004MNRAS.351..505G,Rasia2005ApJ...618L...1R,
  Rasia2008ApJ...674..728R}, to extract mock images and spectra from
simulations and to reduce data using the same procedure followed for
observational data (Rasia et al., in prep., Paper II).  


\section*{Acknowledgements}
 
  We are greatly indebted to Volker Springel for
  providing us with the non--public version of \gadget. We acknowledge
  useful discussions with Gus Evrard, Andrey Kravtsov and Chris
  Miller. This work has been partially supported by PRIN-MIUR grant
  ``The Cosmic Cycle of Baryons'' and PRIN-INAF 2009 grant 
  ``Towards an italian network for computational cosmology'', 
  by ASI--AAE and ASI--COFIS grants and 
  by the INFN--PD51 grant.
  DF acknowledges the support by the European
  Union and Ministry of Higher Education, Science and Technology of
  Slovenia. ER acknowledges the Michigan Society of Fellow. KD
  acknowledges the financial support by the ``HPC-Europa Transnational
  Access program'' and the hospitality of CINECA and of the Department
  of Physics of the University of Trieste. KD also acknowledges the 
  support by the DFG Priority Programme 1177 and additional support by 
  the DFG Cluster of Excellence "Origin and Structure of the Universe". 
  DF, SB and ER acknowledge
  the hospitality of the Sesto Center for Astrophysics (SCfA), where
  part of the work has been carried out. Simulations have been carried
  out at the CINECA Supercomputing Center (Bologna), with CPU time
  assigned thanks to an INAF--CINECA grant and to an agreement between
  CINECA and the University of Trieste.

\bibliographystyle{mn2e}
\bibliography{master}

\end{document}